\documentclass[12pt]{article}
\usepackage{cite,epsfig,amssymb,amsmath,graphicx,color}
\evensidemargin \oddsidemargin
\evensidemargin \oddsidemargin

\newcommand{\nn}{\nonumber}

\begin{document}

\begin{center}
{{{\Large \bf Holography of Massive M2-brane Theory: Non-linear Extension }
}\\[17mm]
O-Kab Kwon$^{1}$,~~Dongmin Jang$^{1}$,
~~Yoonbai Kim$^{1}$,~~D.~D. Tolla$^{1,2}$\\[3mm]
{\it $^{1}$Department of Physics,~BK21 Physics Research Division,
~Institute of Basic Science, Sungkyunkwan University, Suwon 440-746, South Korea\\
$^{2}$University College,\\
Sungkyunkwan University, Suwon 440-746, South Korea}\\[2mm]
{\it okab@skku.edu,~dongmin@skku.edu,~yoonbai@skku.edu,~ddtolla@skku.edu} }

\end{center}
\vspace{15mm}

\begin{abstract}
We investigate the gauge/gravity duality between the ${\cal N} = 6$ mass-deformed ABJM theory with U$_k(N)\times$U$_{-k}(N)$ gauge symmetry and the 11-dimensional supergravity on LLM geometries with SO(2,1)$\times$SO(4)/${\mathbb Z}_k$ $\times$SO(4)/${\mathbb Z}_k$ isometry, in terms of a KK holography, which involves quadratic order field redefinitions.
We establish the quadratic order KK mappings for various gauge invariant fields in order to obtain the canonical 4-dimensional gravity equations of motion and to reduce the LLM solutions to an asymptotically AdS$_4$ gravity solutions. The non-linearity of the KK maps indicates that we can observe the true purpose of the non-linear KK holography of the LLM solutions. Using such KK holography procedure, we obtain the vacuum expectation values of the chiral primary operator of conformal dimension $\Delta = 2$ in the large $N$ limit but with general $k$ and examine the gauge/gravity duality for LLM solutions, which are represented by square-shaped Young diagrams. We also show that the vacuum expectation values of the massive KK graviton modes are vanishing as expected by the supersymmetry.
\end{abstract}

\newpage
\tableofcontents

\section{Introduction}

AdS/CFT correspondence~\cite{Maldacena:1997re,Gubser:1998bc,Witten:1998qj} and its various deformations have been a central paradigm for the past two decades in theoretical physics. 
Among the deformations, we consider the supersymmetry preserving mass deformation~\cite{Hosomichi:2008jb,Gomis:2008vc} of the 3-dimensional ${\cal N} = 6$ U$_k(N)\times {\rm U}_{-k}(N)$ Aharony-Bergman-Jafferis-Maldacena (ABJM) theory with Chern-Simons level $k$~\cite{Aharony:2008ug}, which is dual to the 11-dimensional supergravity on the Lin-Lunin-Maldacena (LLM) geometries~\cite{Lin:2004nb} with ${\mathbb Z}_k$ orbifold and SO(2,1)$\times$SO(4)/${\mathbb Z}_k\times$SO(4)/${\mathbb Z}_k$ isometry. The correspondence between the supersymmetric vacua of the mass-deformed ABJM theory (mABJM) and the LLM geometries with ${\mathbb Z}_k$ orbifold was reported in \cite{Cheon:2011gv}.

Recently, we have disclosed more evidence for the gauge/gravity duality between the mABJM theory and the 11-dimensional supergravity on the LLM geometry with SO(2,1) $\times$SO(4)/${\mathbb Z}_k\times$SO(4)/${\mathbb Z}_k$ isometry~\cite{Jang:2016tbk}. We calculated the vacuum expectation values ($vevs$) of a chiral primary operator (CPO) of conformal dimension $\Delta = 1$, from all supersymmetric vacua of the mABJM theory in the large $N$ limit and from the LLM solutions in the 11-dimensional supergravity in terms of the gauge/gravity dictionary~\cite{Gubser:1998bc,Witten:1998qj}. In order to show the duality, we defined the 4-dimensional dual scalar modes obtained from the procedure of the Kaluza-Klein (KK) holography~\cite{Skenderis:2006uy,Skenderis:2006di,Skenderis:2007yb} for the 11-dimensional supergravity.
We found an exact dual relation between the two results for all possible supersymmetric solutions in both sides in the large $N$ limit.

In the case of the CPO of conformal dimension $\Delta = 1$, linearized Einstein equations and asymptotic expansion of the LLM solutions to the linear order were sufficient to read the $vev$. In that case, the KK maps between the 4-dimensional fields and 11-dimensional fields are trivial. In this paper, we extend to the case of CPO of conformal dimension $\Delta = 2$, which requires non-linear KK maps. We start with the compactification on $S^7/{\mathbb Z}_k$ of the 11-dimensional gravity equations in which the dynamical fields are written as a sum of the AdS$_4\times S^7/{\mathbb Z}_k$ background and fluctuations. To obtain the $vevs$ of the CPO of conformal dimension $\Delta = 2$, it is sufficient to keep up to the quadratic terms in fluctuations. After some manipulations for equations of gauge invariant fluctuation modes, we find that the quadratic terms contain higher derivatives, and thus we need to introduce some non-trivial field redefinitions (the KK maps) to obtain the canonical equations of motion for the 4-dimensional fields. The asymptotically AdS$_4$ solutions to the resulting 4-dimensional equations of motion are obtained from the asymptotic expansion of the LLM solutions and combining various fields in the expansion, according to our well established non-linear KK maps.
Using the holographic renormalization and asymptotic expansion of the LLM geometries, we read the $vevs$ of the CPO of conformal dimension $\Delta = 2$ and also confirm that the $vevs$ of some massive KK graviton modes are vanishing as required by supersymmetry.\footnote{See~\cite{Jang:2017gwd} for results of zeroth KK graviton modes.} On the field theory side, we use the discrete Higgs vacua of the mABJM theory to determine the $vev$ of the CPO of conformal dimension $\Delta=2$ in the large $N$ limit. We check the correspondence of the gravity and the field theory results in the large $N$ limit and general $k$ by considering the case of the LLM geometries represented by square-shaped Young diagrams.

The remaining part of the paper is organized as follows. In section 2, we apply the KK reduction to 11-dimensional supergravity equations and obtain the equations for 4-dimensional gauge invariant fields. We also establish the non-trivial KK maps for some 4-dimensional gauge invariant fields. In section 3, we obtain the CPO of conformal dimension $\Delta = 2$ in the mABJM theory and determine its $vev$ from the discrete Higgs vacua. In section 4, we rearrange the asymptotic expansion of the LLM solutions according to our KK maps to obtain the asymptotically AdS$_4$ solutions of the 4-dimensional gravity equations of motion. From these solutions, we read the $vevs$ of various 4-dimensional KK modes, using the gauge/gravity dictionary. In section 5, we compare the gravity and the field theory results for the $vevs$ of the CPOs and determine the values of some normalization factors. In section 6, we draw our conclusions. In the Appendix, we give some details about the construction of CPO of conformal dimension $\Delta = 2$.

\section{KK Reduction of 11-dimensional Gravity}\label{4dgravity}

In this section, we discuss the compactification of 11-dimensional gravity on $S^7/{\mathbb Z}_k$. The compactification involves expansion of the 11-dimensional fluctuations in terms of the spherical harmonics on $S^7/{\mathbb Z}_k$ and then projecting the equations of motions on those spherical harmonics to obtain the equations of motion for various KK modes. The resulting equations contain higher derivatives of those KK modes and the necessary KK maps are introduced for obtaining the canonical equations of motion of the 4-dimensional dynamical fields.

\subsection{Field equations at quadratic order}
In~\cite{Jang:2017gwd}, we have written the 11-dimensional gravity equations of motion up to quadratic order in the fluctuations by perturbing the fields around the AdS$_4\times S^7/{\mathbb Z}_k$ background as
\begin{align}\label{fluct}
{\bf g}_{pq}=g_{pq}+h_{pq}, \qquad {\bf F}_{pqrs}=F_{pqrs}+f_{pqrs},
\end{align}
where $p,q,\cdots=0,\cdots,10$.
For clarity, we summarize those quadratic order equations. The quadratic order equations are obtained by inserting \eqref{fluct} into the 11-dimensional gravity equations of motion and keeping all the terms up to quadratic order in the fluctuations $h_{pq}$ and $f_{pqrs}$. The results are
\begin{align}
&\nabla^r\nabla_{p}h_{qr}+\nabla^r\nabla_{q}h_{pr}-\nabla^2h_{pq}-\nabla_q\nabla_ph^r{}_{r}-Rh_{pq}-g_{pq}\left(-R^{rs}h_{rs}
+\nabla^r\nabla^sh_{rs}-\nabla^2h^r{}_{r}\right)
\nn\\
&+\frac{1}{48}\Big(F_{rstu}F^{rstu} h_{pq}
{-4}g_{pq}h_{rs}F^r{}_{tuv}F^{stuv}\Big)+\frac{1}{24}g_{pq}f_{rstu}F^{rstu}-\frac12 h_{rs}F^r{}_{ptu}F_q{}^{stu}\nn\\
&-\frac1{6}\Big(
f_{prst}F_q^{~rst}+F_{prst}f_q^{~rst}\Big)+Q_{pq}=0,
\label{lineq2-1}
\\
\label{C3EoM-1}
&
\nabla_p(h^{t}{}_{t}F^{pqrs})+2\nabla_p(4F_{t}^{~[pqr}h^{s]t}+f^{pqrs})
+\frac2{\sqrt{-g}}\frac{1}{(4!)^2}\tilde\epsilon^{p_1\cdots p_4q_1\cdots q_4qrs}f_{p_1\cdots p_4}F_{q_1\cdots q_4}+P^{qrs}=0,
\end{align} 
where the indices are raised (lowered) by the AdS$_4\times S^7/{\mathbb Z}_k$ metric and the covariant derivatives are also those of the background. Here, $P^{qrs}$ and $Q_{pq}$ denote the quadratic terms in the fluctuations and are given by
\begin{align}\label{C3EoM-2}
P^{qrs}=&-\frac12\nabla_p\Big[\Big(h_{tu}h^{tu}-\frac12(h^{t}{}_{t})^2\Big) F^{pqrs}\Big]-8\nabla_p\Big(F_{u}^{~[pqr}h^{s]t}h_t{}^{u}-\frac32F^{tu[pq}~\!\!h^r{}_{t}h^{s]}{}_{u}-f_{t}^{~[pqr}h^{s]t}\Big)\nn\\
&+\nabla_p\Big[h^{t}{}_{t}\big( 4F_{u}^{~[pqr}h^{s]u}+f^{pqrs}\big)\Big]+\frac1{\sqrt{-g}}\frac{1}{(4!)^2}\tilde\epsilon^{p_1\cdots p_4q_1\cdots q_4qrs}f_{p_1\cdots p_4}f_{q_1\cdots q_4},
\end{align} 
\begin{align}\label{lineq1-2}
Q_{pq}=&
-\nabla_r\Big[h^{rs}\big(\nabla_p h_{sq}
+\nabla_q h_{sp}-\nabla_s h_{pq}\big)\Big]+\frac12\nabla_ph^{rs}\nabla_q h_{rs}+h^{rs}\nabla_p\nabla_q h_{rs}\nn\\
&+\frac12\nabla^rh^s{}_s\big(\nabla_p h_{rq}+\nabla_q h_{rp}-\nabla_r h_{pq}\big)
+\nabla^r h^s{}_{p}\nabla_r h_{sq}-\nabla^r h^s{}_{p}\nabla_s h_{qr}-g_{pq}R_{rs}h^{rt}h^s{}_t\nn\\
&+\frac12g_{pq}\nabla_r\Big[h^{rs}\big(2\nabla^t h_{st}-\nabla_s h^t{}_t\big)\Big]-\frac34g_{pq}\nabla^th^{rs}\nabla_t h_{rs}+\frac12g_{pq}\nabla^r h^{st}\nabla_s h_{tr}-\frac12g_{pq}h^{rs}\nabla^2 h_{rs}\nn\\ 
&-\frac14g_{pq}\nabla^rh^s{}_s\big(2\nabla^t h_{rt}-\nabla_r h^t{}_t\big)
+\frac12g_{pq}h^{rs}\Big(\nabla^t\nabla_rh_{ts}+\nabla^t\nabla_sh_{tr}-\nabla^2h_{rs}-\nabla_r\nabla_sh^t{}_t\Big)\nn\\
&+h_{pq}h^{rs}R_{rs}-h_{pq}\Big(\nabla^r\nabla^sh_{rs}-\nabla^2h^r{}_r\Big)+\frac1{12}\Big[g_{pq}F_{rstu}F^{rst}{}_{w}h^{uv}h_v{}^{w}-g_{pq}F_{rstu}f^{rst}{}_{v}h^{uv}\nn\\
&+\frac32g_{pq}F_{rstu}F^{rs}{}_{vw}h^{tv}h^{uw}+\frac12h_{pq}f_{rstu}F^{rstu}-h_{pq}F_{rstu}F^{rst}{}_{v}h^{uv}+\frac14g_{pq}f_{rstu}f^{rstu}\nn\\
&-g_{pq}f_{rstu}F^{rst}{}_{v}h^{uv}\Big]-\frac1{2}\Big[F_{pstu}F_q{}^{st}{}_{w}h^{uv}h_v{}^{w}+F_{pstu}F_q{}^{s}{}_{vw}h^{tv}h^{uw}-F_{pstu}f_q{}^{st}{}_{v}h^{uv}\nn\\
&-f_{pstu}F_q{}^{st}{}_{v}h^{uv}+\frac13f_{pstu}f_q{}^{stu}\Big].
\end{align}

The KK reduction of the 11-dimensional gravity to 4-dimensional gravity involves the expansion of the fluctuations $h_{pq}$ and $f_{pqrs}$ in terms of the spherical harmonics on $S^7/{\mathbb Z}_k$, with the metric
\begin{align}
 ds_{S^7/{\mathbb Z}_k}^2=d\tau^2+ ds_{S^3/{\mathbb Z}_k}^2 + ds_{\tilde S^3/{\mathbb Z}_k}^2.
\end{align}
Later, we will identify the fluctuations $h_{pq}$ and $f_{pqrs}$ with the deviations of the LLM solutions from the AdS$_4\times S^7/{\mathbb Z}_k$ solutions. Keeping in mind the SO(2,1)$\times {\rm SO}(4)/{\mathbb Z}_k \times {\rm SO}(4)/{\mathbb Z}_k$ isometry of the LLM solutions, we consider expansions in terms of the spherical harmonics with $ {\rm SO}(4)/{\mathbb Z}_k \times {\rm SO}(4)/{\mathbb Z}_k$ symmetry. Since those spherical harmonics depend only on the $\tau$ coordinate, they are not affected by the orbifolding. This implies that expansions of the fluctuations $h_{pq}$ and $f_{pqrs}$ in terms of these spherical harmonics are the same, irrespective of the orbifolding.
In~\cite{Jang:2016tbk}, we have written a complete form of these expansions whereas we have argued in~\cite{Jang:2017gwd} that many of the KK modes do not contribute to the equations of motion in quadratic order. Therefore, we use the following truncated expansions,
\begin{align}\label{metric-exp1}
&h_{\mu\nu}(x,y)=h^{I_1}_{\mu\nu}(x)Y^{I_1}(y),\quad h^\rho{}_{\rho}(x,y)=h^{I_1}(x)Y^{I_1}(y),\nn\\
&h_{(ab)}=s^{I_1}(x)\nabla_{(a}\nabla_{b)}Y^{I_1}(y),\quad h^a_{~a}(x,y)=\phi^{I_1}(x)Y^{I_1}(y),\nn\\
&f_{\mu\nu\rho\sigma}(x,y)=\frac23\nabla_{[\mu} t^{\lambda I_1}(x)\epsilon_{\nu\rho\sigma]\lambda}Y^{I_1}(y),\quad f_{\mu\nu\rho a}(x,y)=- \frac1{3!}\epsilon_{\mu\nu\rho}\!\!~^{\sigma}t_\sigma^{I_1}(x)\nabla_{a}Y^{I_1}(y),\nn\\
&f_{\mu abc}(x,y)=\nabla_\mu t^{I_{35}}(x)Y_{abc}^{I_{35}}(y)
,\quad f_{abcd}(x,y)=4 t^{I_{35}}(x)\nabla_{[a}Y_{bcd]}^{I_{35}}(y),
\end{align}
where $I_{n}=0,1,2,\cdots$, we have split the 11-dimensional indices into the AdS$_4$ indices
$(\mu,\nu,\cdots =0,\cdots,3)$ and the $S^7$ indices $(a,b,\cdots=4,\cdots,10)$, $x$ denotes the AdS${}_4$ coordinates and $y$ denotes the $S^7$ coordinates. The notation $(ab)$ means symmetrized traceless combination, while $[ab\cdots]$ denotes complete antisymmetrization of indices. Here, $Y^{I_{1}}$ and $Y_{abc}^{I_{35}}$ are the scalar and antisymmetric 3-tensor spherical harmonics on $S^{7}$, respectively.

Plugging \eqref{metric-exp1} into the ($\mu\nu)$ component of \eqref{lineq2-1} and then projecting on the scalar spherical harmonics $Y^{I_1}$, we obtain
\begin{align}\label{lineqmn}
&-\left(\square+\Lambda^{I_1}-\frac{24}{L^2}\right) h^{I_1}_{\mu\nu}+\nabla^\rho\nabla_{\mu}h^{I_1}_{\nu\rho}+\nabla^\rho\nabla_{\nu}h^{I_1}_{\mu\rho}
-\nabla_\mu\nabla_\nu (h^{I_1}+\phi^{I_1})\nn\\
&+g_{\mu\nu}\left(\square+\Lambda^{I_1}-\frac{30}{L^2}\right)h^{I_1}
+g_{\mu\nu}\left(\square+\frac67\Lambda^{I_1}+\frac6{L^2}\right)\phi^{I_1}-g_{\mu\nu}\left(\frac67\Lambda^{I_1}+\frac6{L^2}\right)\Lambda^{I_1}s^{I_1}\nn\\
&-g_{\mu\nu}\nabla^\rho\nabla^\sigma h^{I_1}_{\rho\sigma}+\frac{1}{L}g_{\mu\nu}\nabla^{\rho} t^{I_1}_\rho+2Q^{I_1}_{\mu\nu}=0,
\end{align}
 where $\square\equiv\nabla_\mu\nabla^\mu$, $L$ is the radius of $S^7$, ~$Q^{I_1}_{\mu\nu}=\frac1{\omega_7}\int_{S^7}\frac12Q_{\mu\nu}Y^{I_1}$, and $\Lambda^{I_{1}}=-\frac{I_{1}(I_{1}+6)}{L^{2}}$ is the eigenvalue corresponding to the scalar harmonics $Y^{I_1}$. The trace of the above equation leads to
\begin{align}\label{h-eqn}
&\Big(\square+\frac32\Lambda^{I_1}-\frac{48}{L^2}\Big)h^{I_1}-\nabla^\mu\nabla^\nu h_{\mu\nu}+\frac32\Big(\square+\frac87\Lambda^{I_1}+\frac8{L^2}\Big)\phi^{I_1}+\frac2{L}\nabla^{\rho} t^{I_1}_\rho\nn\\
&-12\Lambda^{I_1}\Big(\frac17\Lambda^{I_1}
+\frac1{L^2}\Big)s^{I_1}+Q_h^{I_1}=0,
\end{align}
where $Q_h^{I_1}=g^{\mu\nu}Q_{\mu\nu}^{I_1}$. 
Secondly, projecting the ($\mu a$) component of \eqref{lineq2-1} on $\nabla^a Y^{I_1} ( I_1\ne0)$\footnote{See~\cite{Jang:2017gwd} for the zeroth mode results.}, we obtain
\begin{align}\label{leq2-2-2}
&-\Big(\frac67 \Lambda^{I_1}+ \frac6{L^2}\Big)\nabla_\mu s^{I_1}+\frac67\nabla_\mu\phi^{I_1}
-\nabla^\nu h_{\mu\nu}^{I_1}+\nabla_\mu h^{I_1}
-\frac1{L} t^{I_1}_\mu+Q_{\mu}^{I_1}
=0,
\end{align}
where $Q_{\mu}^{I_1}=\frac1{\omega_7}\int_{S^7}Q_{\mu a}\nabla^aY^{I_1}$.
Thirdly, projecting the $(ab)$ component of \eqref{lineq2-1} on $ g^{ab} Y^{I_1}$ and $\nabla^{(a}\nabla^{b)} Y^{I_1}(I_1\ne 0)$, we obtain two scalar equations
 \begin{align}
&
3\Big(\square+\frac{5}7\Lambda^{I_1}+\frac{5}{L^2}\Big)\phi^{I_1}
+\frac72\Big(\square+\frac67\Lambda^{I_1}+\frac{6}{L^2}\Big)h^{I_1}
-\frac72\nabla^\mu\nabla^\nu h_{\mu\nu}^{I_1}
-\frac7{2L}\nabla^{\rho} t^{I_1}_\rho\nn\\
&
-15\Lambda^{I_1}\Big(\frac{\Lambda^{I_1}}7+\frac1{L^2}
\Big)s^{I_1}+Q_\phi^{I_1}=0,\label{leq3-4-2}\\
&\Lambda^{I_1}\Bigg\{\Big(\square-\frac57\Lambda^{I_1}\Big)s^{I_1}+h^{I_1}+\frac{5}{7}\phi^{I_1}\Bigg\}-Q_s^{I_1}=0
,\label{leq3-3-2}
\end{align}
where $\ Q_{\phi}^{I_1}=\frac1{\omega_7}\int_{S^7}\frac 12Q_{ab}g^{ab}Y^{I_1}$ and $ Q^{I_1}_s=\frac1{7}\Big(6\Lambda^{I_1}+\frac{42}{L^2}\Big)^{-1}\frac1{\omega_7}\int_{S^7}{Q_{ab}}\nabla^{(a}\nabla^{b)} Y^{I_1}$.
Similarly, inserting \eqref{metric-exp1} into $(\mu\nu\rho)$ component of \eqref{C3EoM-2} and projecting on $Y^{I_1}$, we obtain the following equation\footnote{More equations can be obtained by projecting the $(\mu\nu a, \mu ab, abc)$ components of \eqref{C3EoM-2} on appropriate spherical harmonic elements, however those equations are not required for our purpose here. See~\cite{Jang:2017gwd} for the full list of equations.}
\begin{align}
&\frac23\nabla^{\sigma}\nabla_{[\sigma} t^{\lambda I_1}\epsilon_{\mu\nu\rho]\lambda}+\frac{\Lambda^{I_1}}{3!}\epsilon_{\mu\nu\rho}\!\!~^{\sigma}t_\sigma^{I_1}-\frac3{L}\epsilon_{\sigma\mu\nu\rho}\nabla^\sigma\big( h^{I_1}+\phi^{I_1}\big)-\frac{24}{L}\nabla^\sigma h^{I_1}_{\lambda[\sigma}\epsilon_{\mu\nu\rho]}~\!\!^{\lambda}+P_{\mu\nu\rho}^{I_1}=0,\label{tSmnr1}
\end{align}
where $P_{\mu\nu\rho}^{I_1}=\frac1{\omega_7}\int_{S^7}P_{\mu\nu\rho}Y^{I_1}$. Applying $\epsilon_{\mu'}^{\!~~\mu\nu\rho}\nabla_{\nu'}$ to \eqref{tSmnr1}, we obtain
\begin{align}\label{C3EoM-4b}
&-\frac{18}{ \Lambda^{I_1}}\nabla_{\mu}\nabla_{\nu}(-h^{I_1}+\phi^{I_1})-L\nabla_\nu t^{I_1}_\mu-\frac L{\Lambda^{I_1}}\nabla_{\mu}\nabla_{\nu}\nabla^\rho t^{I_1}_\rho+\tilde Q^{I_1}_{\mu\nu}=0,
\end{align}
where $\tilde Q^{I_1}_{\mu\nu}=-\frac L{\Lambda^{I_1}}\epsilon_{\mu}^{\!~~\rho\sigma\lambda}\nabla_{\nu}P^{I_1}_{\rho\sigma\lambda}$. The trace of the above equation gives
\begin{align}\label{C3EoM-4c}
&\frac{18}L\square(-h^{I_1}+\phi^{I_1})+(\square+\Lambda^{I_1})\nabla^\rho t^{I_1}_\rho+Q^{I_1}_{\psi}=0,
\end{align}
where $Q^{I_1}_{\psi}=-\Lambda^{I_1}g^{\mu\nu}\tilde Q^{I_1}_{\mu\nu}$.

\subsection{Quadratic order equations for KK modes}
The quadratic order equations we listed in the previous subsection lead to the quadratic order equations of motions for various 4-dimensional gauge invariant KK modes. In general, the 4-dimensional gravity spectrum, which is obtained from the KK reduction of the 11-dimensional gravity, is composed of three towers of scalar modes, two towers of pseudoscalar modes, two towers of vector modes, one tower of pseudovector modes, and one tower of spin-two modes~\cite{Jang:2016tbk}. Here, we follow the gauge choice of the LLM solutions in which $h_{\mu a}$ and $f_{\mu\nu ab}$ are zero and as a result some of the KK towers are absent. In addition, in this paper, we are interested in the gravity field which is dual to the CPO of conformal dimension $\Delta = 2$ in the mABJM theory. Such dual gravity field is a part of the three KK towers of scalar modes with $I_1=4$. Therefore, from now on we focus on the equations of motion for the KK modes with $I_1=4$. 

Setting $I_1=4$ in \eqref{lineqmn} $-$ \eqref{C3EoM-4c} and rearranging the equations, we obtain the following set of equations, 
\begin{align}\label{hhmnEq}
&\square\hat h^4_{\mu\nu}=\frac{32}{L^2}\hat h^4_{\mu\nu}+\frac1{20}\nabla_\mu\nabla_\nu\hat\psi^4-\frac9{10}\nabla_\mu\nabla_\nu\hat\phi^4-\frac4{3L^2}g_{\mu\nu}\hat\psi^4-\frac{40}{7L^2}g_{\mu\nu}\hat\phi^4+\nabla_\mu Q^4_\nu+\nabla_\nu Q^4_\mu\nn\\
&~~~~~~~+\frac{L^2}{40}\nabla_\mu\nabla_\nu Q^4_s-\frac29g_{\mu\nu}\Big(Q^4_h+Q^4_\phi-\frac9{10}Q^4_s\Big)-\frac1{L^2}(\tilde Q^4_{\mu\nu}+\tilde Q^4_{\nu\mu})+2Q^4_{\mu\nu},\\
& \hat J^4_{\mu\nu}=-\frac{L^2}{40}\nabla_{\mu}\nabla_{\nu}\hat\psi^4+\frac{9L^2}{20}\nabla_{\mu}\nabla_{\nu}\hat\phi^4+\frac12(\tilde Q^4_{\mu\nu}+\tilde Q^4_{\nu\mu}),\label{UmnEq3}\\
&\square\hat\phi^4=\frac{28}{L^2}\hat\phi^4+\frac{14}{3L^2}\hat\psi^4-\frac{14}9Q^4_h+\frac49Q^4_\phi-Q^4_s ,\label{hatphi}\\
&\square\hat\psi^4=\frac{124}{L^2}\hat\psi^4+\frac{7128}{7L^2}\hat\phi^4-28Q^4_h+8Q^4_\phi+Q^4_\psi,\label{hatpsi}
\end{align}
where we have introduced $u^{I_1}_{\mu\nu}\equiv\frac L2(\nabla_\mu t^{I_1}_\nu+\nabla_\nu t^{I_1}_\mu),$ $u^{I_1}\equiv g^{\mu\nu}u^{I_1}_{\mu\nu}$, and the following gauge invariant combinations, 
\begin{align}
&\hat h^{I_1}_{\mu\nu}\equiv h^{I_1}_{\mu\nu}+\nabla_\mu\nabla_\nu S^{I_1},\qquad\,\hat J^{I_1}_{\mu\nu}\equiv u^{I_1}_{\mu\nu}+18\nabla_\mu\nabla_\nu S^{I_1}, \nn\\
& \hat\phi^{I_1}\equiv \phi^{I_1}-\Lambda^{I_1}s^{I_1},\qquad\qquad\hat\psi^{I_1}\equiv 18 h^{I_1}-u^{I_1}.
\end{align}

\subsubsection{Spin-zero field equations}
The equations of motion for spin-zero mass eigenstates are given by the linear combinations of \eqref{hatphi} and \eqref{hatpsi}. Introducing the mass eigenstates 
\begin{align}
\check\phi^4=\frac{297}{49}\hat\phi^4+\frac{11}{14}\hat\psi^4,\qquad \check\psi^4=-\frac{297}{49}\hat\phi^4+\frac{3}{14}\hat\psi^4,\end{align}
and combining \eqref{hatphi} and \eqref{hatpsi}, we obtain the following diagonalized equations 
\begin{align}\label{diag-scalar}
&\Big(\square+\frac8{L^2}\Big)\check\psi^4-\frac{24}7Q^4_h+\frac{48}{49}Q^4_\phi-\frac3{14}Q^4_\psi+\frac{297}{49}Q^4_s=0,\nn\\
&\Big(\square-\frac{160}{L^2}\Big)\check\phi^4+\frac{220}7Q^4_h-\frac{440}{49}Q^4_\phi-\frac{11}{14}Q^4_\psi-\frac{297}{49}Q^4_s=0.
\end{align}
All the quadratic terms in the above equations are composed of the expressions which are quadratic in the fields $h^{I_1}_{\mu\nu},t^{I_1}_{\mu}, h^{I_1},\phi^{I_1}, u^{I_1}, s^{I_1}, t^{I_{35}}$ and their derivatives, with infinite summations over $I_{1}$ and $I_{35}$. The LLM solution solves the 11-dimensional equations of motion order by order in the mass parameter $\mu_0$ of the LLM geometries~\cite{Jang:2016tbk,Jang:2017gwd}. In the above equations of motion, we have kept only up to the quadratic terms in the fluctuations and they are expected to be solved by the LLM solution only up to quadratic order in $\mu_0$. On the other hand, except for the modes with ${I_1}=2$ and $I_{35}=1$, the asymptotic expansions of the other modes are non-linear in the expansion parameter $\mu_0$. Thus, the relevant quadratic terms in the above equations are built only by the modes with ${I_1}=2$ and $I_{35}=1$. In addition, we note that for the spherical harmonics on $S^7$ with ${\rm SO}(4)\times {\rm SO}(4)$ symmetry, (See \cite{Jang:2016tbk})
\begin{align}
\int_{S^7}Y^4Y^1_{abc}~g^{aa'}g^{bb'}g^{cc'}Y^1_{a'b'c'}=0,\quad \int_{S^7}\nabla^a\nabla^bY^4Y^1_{acd}~g^{cc'}g^{dd'}Y^1_{bc'd'}=0.\end{align}
The LLM solutions depend only on such spherical harmonics. In that case, the terms involving $ t^{I_{35}=1}$ are also absent and the quadratic terms depend only on $h^{2}_{\mu\nu},t^{2}_{\mu}, h^{2},\phi^{2}, u^{2}, s^{2}$ and their derivatives. Combining the four scalar fields $h^{2},\phi^{2}, u^{2}, s^{2}$, we obtain two gauge invariant physical mode, $\check\phi^2=\frac9{70}(7\hat\psi^2+18\hat\phi^2),~\check\psi^2=\frac1{70}(7\hat\psi^2-162\hat\phi^2)$, which are mass eigenstates. The other potentially relevant gauge invariant physical mode is the second KK graviton mode, which is given by 
\begin{align}
\check h_{(\mu\nu)}^{2}=\hat h^{2}_{(\mu\nu)}
-\frac1{4}\hat J^{2}_{(\mu\nu)}
+\frac{15L^2}{112}\nabla_{(\mu}\nabla_{\nu)}\hat\phi^{2}
-\frac{L^2}{96}\nabla_{(\mu}\nabla_{\nu)}\hat\psi^{2}. 
\end{align}

In general, our quadratic terms depend on the two physical scalar modes ($\check\phi^2,\check\psi^2$) and the second KK graviton mode $\check h_{(\mu\nu)}^{2}$. However, the leading order terms in the asymptotic expansions of $\check h_{(\mu\nu)}^{2}$ and $\check\phi^2$ are $\mu_0^{3}$-order, and they are irrelevant for quadratic order equations. As a result, the otherwise very complex quadratic terms are composed of only $\check\psi^2$, and are given by
\begingroup
\allowdisplaybreaks
\begin{align}\label{Quad1}
Q^4_h=&-\frac{1}{41472\sqrt{10}L^{2}}\bigg(41216\check\psi^2\check\psi^2+2560L^{2}\nabla_\rho\check\psi^2\nabla^\rho\check\psi^2+88L^{4}\nabla_\rho\nabla_\sigma\check\psi^2\nabla^\rho\nabla^\sigma\check\psi^2
\nn\\
&~~~~~~~~~~~~~~~~~~~~~~~-\nabla_\rho\nabla_\sigma\nabla_\lambda\check\psi^2\nabla^\rho\nabla^\sigma\nabla^\lambda\check\psi^2\bigg),
\nn\\
Q^4_\phi=&-\frac{1}{82944\sqrt{10}L^{2}}\bigg(126080\check\psi^2\check\psi^2+12736L^{2}\nabla_\rho\check\psi^2\nabla^\rho\check\psi^2-32L^{4}\nabla_\rho\nabla_\sigma\check\psi^2\nabla^\rho\nabla^\sigma\check\psi^2
\nn\\
&~~~~~~~~~~~~~~~~~~~~~~~-7L^{6}\nabla_\rho\nabla_\sigma\nabla_\lambda\check\psi^2\nabla^\rho\nabla^\sigma\nabla^\lambda\check\psi^2\bigg),
\nn\\
Q^4_s=&-\frac{5}{1944\sqrt{10}L^2}\bigg(120\check\psi^2\check\psi^2+8L^{2}\nabla_\rho\check\psi^2\nabla^\rho\check\psi^2+L^{4}\nabla_\rho\nabla_\sigma\check\psi^2\nabla^\rho\nabla^\sigma\check\psi^2\bigg),
\nn\\
Q^4_\psi=&-\frac{1}{576\sqrt{10}L^2}\bigg(3584\check\psi^2\check\psi^2+416L^{2}\nabla_\rho\check\psi^2\nabla^\rho\check\psi^2-80L^{4}\nabla_\rho\nabla_\sigma\check\psi^2\nabla^\rho\nabla^\sigma\check\psi^2\bigg)
\nn\\
&~~~~~~~~~~~~~~~~~~~~+L^{6}\nabla_\rho\nabla_\sigma\nabla_\lambda\check\psi^2\nabla^\rho\nabla^\sigma\nabla^\lambda\check\psi^2,\nn\\
Q^4_{\mu\nu}=&
-\frac{1}{41472\sqrt{10}L^2}\bigg[\frac{1}{2}g_{\mu\nu}\bigg(22400\check\psi^2\check\psi^2
+1600L^{2}\nabla_\rho\check\psi^2\nabla^\rho\check\psi^2
+48L^{4}\nabla_\rho\nabla_\sigma\check\psi^2\nabla^\rho\nabla^\sigma\check\psi^2
\nn\\
&~~~~~~~~~~~~~~~~~~~~~~~~~~~~~~
-L^4\nabla_\rho\nabla_\sigma\nabla_\lambda\check\psi^2\nabla^\rho\nabla^\sigma\nabla^\lambda\check\psi^2\bigg)
\nn\\
&~~~~~~~~~~~~~~~~~~~~~
+320\nabla_\mu\check\psi^2\nabla_\nu\check\psi^2
+448L^{2}\check\psi^2\nabla_\mu\nabla_\nu\check\psi^2
+48L^{4}\nabla_\rho\check\psi^2\nabla_\mu\nabla_\nu\nabla^\rho\check\psi^2
\nn\\
&~~~~~~~~~~~~~~~~~~~~~
-8L^{6}\nabla_\mu\nabla_\rho\check\psi^2\nabla_\nu\nabla^\rho\check\psi^2
+L^{6}\nabla_\mu\nabla_\rho\nabla_\sigma\check\psi^2\nabla_\nu\nabla^\rho\nabla^\sigma\check\psi^2\bigg],
\nn\\
Q^4_\mu=&\frac{1}{41472\sqrt{10}}\bigg(1568\check\psi^2\nabla_\mu\check\psi^2
+24L^2\nabla_\rho\check\psi^2\nabla_\mu\nabla^\rho\check\psi^2
+L^4\nabla_\rho\nabla_\sigma\check\psi^2\nabla_\mu\nabla^\rho\nabla^\sigma\check\psi^2\bigg),
\nn\\
\tilde Q^4_{\mu\nu}=&\frac{L^2}{23040\sqrt{10}}\bigg(384\nabla_\mu\check\psi^2\nabla_\nu\check\psi^2
+40L^{2}\nabla_\mu\nabla_\rho\check\psi^2\nabla_\nu\nabla^\rho\check\psi^2
-L^4\nabla_\mu\nabla_\rho\nabla_\sigma\check\psi^2\nabla_\nu\nabla^\rho\nabla^\sigma\check\psi^2
\nn \\
&~~~~~~~~~~~~~~~
+384\check\psi^2\nabla_\mu\nabla_\nu\check\psi^2
+40L^{2}\nabla_\rho\check\psi^2\nabla_\mu\nabla_\nu\nabla^\rho\check\psi^2
-L^4\nabla_\rho\nabla_\sigma\check\psi^2\nabla_\mu\nabla_\nu\nabla^\rho\nabla^\sigma\check\psi^2\bigg).
\end{align}
\endgroup
Inserting these quadratic terms into \eqref{diag-scalar}, we obtain
\begin{align}\label{11DEq}
&\Big(\square+\frac8{L^2}\Big)\check\psi^4+\frac{1}{8064\sqrt{10}L^2}\Big(11136\check\psi^2\check\psi^2+736L^2\nabla_\rho\check\psi^2\nabla^\rho\check\psi^2-304L^4\nabla_\rho\nabla_\sigma\check\psi^2\nabla^\rho\nabla^\sigma\check\psi^2
\nn\\
&~~~~~~~~~~~~~~~~+3L^6\nabla_\rho\nabla_\sigma\nabla_\lambda\check\psi^2\nabla^\rho\nabla^\sigma\nabla^\lambda\check\psi^2\Big)=0,
\nn\\
&\Big(\square-\frac{160}{L^2}\Big)\check\phi^4+\frac{11}{8064\sqrt{10}L^2}\Big(-7936\check\psi^2\check\psi^2+96L^2\nabla_\rho\check\psi^2\nabla^\rho\check\psi^2-120L^4\nabla_\rho\nabla_\sigma\check\psi^2\nabla^\rho\nabla^\sigma\check\psi^2
\nn\\
&~~~~~~~~~~~~~~~~~~+L^6\nabla_\rho\nabla_\sigma\nabla_\lambda\check\psi^2\nabla^\rho\nabla^\sigma\nabla^\lambda\check\psi^2\Big)=0.
\end{align}
This shows that the usual compactification of the 11-dimensional supergravity on $S^7$ results in the field equations which contain higher derivative terms. In order to obtain the canonical 4-dimensional gravity equations of motion, we need to introduce some field redefinitions to absorb those higher derivative 
terms~\cite{Lee:1998bxa,Arutyunov:1999en,Skenderis:2006uy,Jang:2017gwd}. The 4-dimensional gravity equations of motion should read as follows,
\begin{align}\label{4DEq}
\Big(\square+\frac8{L^2}\Big)\Psi^4+\alpha\Psi^2\Psi^2=0,\qquad\Big(\square-\frac{160}{L^2}\Big)\Phi^4+\beta\Psi^2\Psi^2=0,
\end{align}
where $\Psi^2\equiv\check\psi^2$.
Since the equations in \eqref{11DEq} contain the terms with up to sextic derivatives, the field redefinitions absorbing those sextic derivatives should contain terms with up to quartic derivatives
\begin{align}\label{fieldRed}
& \Psi^4=\check\psi^4+A_1\check\psi^2\check\psi^2+A_2\nabla_\rho\check\psi^2\nabla^\rho\check\psi^2+A_3\nabla_\rho\nabla_\sigma\check\psi^2\nabla^\rho\nabla^\sigma\check\psi^2, 
\nn\\
&\Phi^4=\check\phi^4+B_1\check\psi^2\check\psi^2+B_2\nabla_\rho\check\psi^2\nabla^\rho\check\psi^2+B_3\nabla_\rho\nabla_\sigma\check\psi^2\nabla^\rho\nabla^\sigma\check\psi^2.
\end{align}
Insertion of \eqref{fieldRed} into \eqref{4DEq} and comparison with \eqref{11DEq} fix the unknown coefficients in \eqref{4DEq} $-$ \eqref{fieldRed} as
\begin{align}
&A_1=-\frac{25}{168\sqrt{10}},\qquad A_2=-\frac{7L^2}{576\sqrt{10}},\qquad A_3=\frac{L^4}{5376\sqrt{10}},\qquad \alpha=0,\nn\\
 &B_1=\frac{11}{168\sqrt{10}},\qquad B_2=0,\qquad B_3=\frac{11L^4}{5376\sqrt{10}},\qquad\beta=0.
\end{align}
The field redefinition of the type \eqref{fieldRed} is usually called the KK map between the 11-dimensional fields ($\check\psi^4,\check\phi^4$) and the 4-dimensional fields ($\Psi^4,\Phi^4$).
\subsubsection{Spin-two field equations}
The equation of motion for the fourth KK graviton mode is a linear combinations of the equations \eqref{hhmnEq}-\eqref{hatpsi}. Let us define the spin-two mass eigenstate as
\begin{align}\label{hhmn}
\check h_{\mu\nu}^{4}&=\hat h^{4}_{\mu\nu}+a_1\hat J^4_{\mu\nu}+a_2\nabla_{\mu}\nabla_{\nu}\hat\phi^{4}+a_3\nabla_{\mu}\nabla_{\nu}\hat\psi^{4}+g_{\mu\nu}(c\hat\phi^{4}+d\hat\psi^{4})\nn\\
&=\hat h^{4}_{\mu\nu}+a\nabla_{\mu}\nabla_{\nu}\hat\phi^{4}+b\nabla_{\mu}\nabla_{\nu}\hat\psi^{4}+g_{\mu\nu}(c\hat\phi^{4}+d\hat\psi^{4}),
\end{align}
where in the second line, we have used the algebraic equation \eqref{UmnEq3} to eliminate $\hat J^4_{\mu\nu}$ up to a redundant quadratic term, which we omit from the definition. Organizing the equations \eqref{hhmnEq} $-$ \eqref{hatpsi} according to this definition and setting
\begin{align}
a=-\frac{17L^2}{1120},\qquad b=\frac{L^2}{2880},\qquad c=\frac{11}{56},\qquad d=\frac1{144},
\end{align}
we obtain the diagonalized equation for the mass eigenstate
\begin{align}\label{Eqhhmn}
\Big(\square-\frac{32}{L^2}\Big)\check h^4_{\mu\nu}&-\bigg[\nabla_\mu Q^4_\nu+\nabla_\nu Q^4_\mu+\frac{L^2}{72}\nabla_\mu\nabla_\nu \Big(Q^4_h+\frac{99}{140} Q^4_s-\frac{2}{7} Q^4_\phi+\frac{1}{40}Q^4_\psi\Big)\nn\\
&-g_{\mu\nu}\Big(\frac{11}{18} Q^4_h-\frac{11}{40}Q^4_s+\frac{1}{9} Q^4_\phi-\frac{7}{720} Q^4_\psi\Big)-\frac2{L^2}\tilde Q^4_{\mu\nu}+2 Q^4_{\mu\nu}\bigg]=0.\end{align}
Inserting the quadratic terms in \eqref{Quad1} into this equation, we rewrite \eqref{Eqhhmn} as
\begin{align}\label{Eqhmn}
\Big(\square&-\frac{32}{L^2}\Big)\check h_{\mu\nu}^{4}-\frac{1}{1080\sqrt{10}} \bigg[2g_{\mu\nu}\Big(\frac{49}{L^2}\check\psi^2\check\psi^2+\frac{137}{72}\nabla_\rho\check\psi^2\nabla^\rho\check\psi^2+\frac{115L^2}{288}\nabla_\rho\nabla_\sigma\check\psi^2\nabla^\rho\nabla^\sigma\check\psi^2\nn\\
&-\frac{7L^4}{768}\nabla_\rho\nabla_\sigma\nabla_\lambda\check\psi^2\nabla^\rho\nabla^\sigma\nabla^\lambda\check\psi^2\Big)+\nabla_\mu\check\psi^2\nabla_\nu\check\psi^2+\frac{17}{3}\check\psi^2\nabla_\mu\nabla_\nu\check\psi^2\nn\\
&-\frac{259L^2}{72}\nabla_\mu\nabla_\rho\check\psi^2\nabla_\nu\nabla^\rho\check\psi^2-\frac{469L^2}{72}\nabla_\rho\check\psi^2\nabla_\mu\nabla_\nu\nabla^\rho\check\psi^2\nn\\
&+\frac{11L^4}{144}\nabla_\mu\nabla_\rho\nabla_\sigma\check\psi^2\nabla_\nu\nabla^\rho\nabla^\sigma\check\psi^2+\frac{37L^4}{288}\nabla_\rho\nabla_\sigma\check\psi^2\nabla_\mu\nabla_\nu\nabla^\rho\nabla^\sigma\check\psi^2\nn\\
&-\frac{L^6}{768}\nabla_\mu\nabla_\rho\nabla_\sigma\nabla_\lambda\check\psi^2\nabla_\nu\nabla^\rho\nabla^\sigma\nabla^\lambda\check\psi^2-\frac{L^6}{768}\nabla_\rho\nabla_\sigma\nabla_\lambda\check\psi^2\nabla_\mu\nabla_\nu\nabla^\rho\nabla^\sigma\nabla^\lambda\check\psi^2\bigg]=0.
\end{align}
This spin-two field equation contains the terms with up to octic derivatives. In order to absorb these higher derivative terms, we need to introduce another field redefinition with up to sextic derivatives as follows
\begin{align}\label{fredhmn}
H^4_{\mu\nu}&=\check h_{\mu\nu}^{4}+g_{\mu\nu}\Big(\tilde C_0\check\psi^2\check\psi^2+\tilde C_1\nabla_\rho\check\psi^2\nabla^\rho\check\psi^2+\tilde C_2\nabla_\rho\nabla_\sigma\check\psi^2\nabla^\rho\nabla^\sigma\check\psi^2+\tilde C_3\nabla_\rho\nabla_\sigma\nabla_\lambda\check\psi^2\nabla^\rho\nabla^\sigma\nabla^\lambda\check\psi^2\Big)\nn\\
&+C_1\nabla_{\mu}\check\psi^2\nabla_{\nu}\check\psi^2++D_1\check\psi^2\nabla_{\mu}\nabla_{\nu}\check\psi^2+C_2\nabla_{\mu}\nabla^\rho\check\psi^2\nabla_{\nu}\nabla_\rho\check\psi^2+D_2\nabla^\rho\check\psi^2\nabla_{\mu}\nabla_{\nu}\nabla_\rho\check\psi^2\nn\\
&+C_3\nabla_{\mu}\nabla^\rho\nabla^\sigma\check\psi^2\nabla_{\nu}\nabla_\rho\nabla_\sigma\check\psi^2+D_3\nabla^\rho\nabla^\sigma\check\psi^2\nabla_{\mu}\nabla_{\nu}\nabla_\rho\nabla_\sigma\check\psi^2.
\end{align} 
 Then the equation of motion of the spin-two field $H^4_{\mu\nu}$ should read 
 \begin{align}\label{EqHmn}
\Big(\square-\frac{32}{L^2}\Big)H^4_{\mu\nu}+g_{\mu\nu}\big(\alpha_0 \Psi^2\Psi^2+\alpha_1\nabla_\rho\Psi^2\nabla^\rho\Psi^2\big)+\beta_1\nabla_{\mu}\Psi^2\nabla_{\nu}\Psi^2+\beta_2\Psi^2\nabla_{\mu}\nabla_{\nu}\Psi^2=0.\end{align}
Inserting \eqref{fredhmn} into \eqref{EqHmn} and comparing with \eqref{Eqhmn}, we determine the unknown coefficients as
\begin{align}
&D_3=\frac{L^6}{1658880\sqrt{10}},\quad C_3=\frac{L^6}{1658880\sqrt{10}},\quad \tilde C_3=0,\quad D_2=-\frac{L^4}{62208\sqrt{10}},\nn\\
&C_2=\frac{L^4}{124416\sqrt{10}},\quad \tilde C_2=\frac{L^4}{165888\sqrt{10}},\quad D_1=\frac{103L^2}{51840\sqrt{10}},\nn\\
&C_1=\frac{113L^2}{51840\sqrt{10}},\quad \tilde C_1=\frac{L^2}{7776\sqrt{10}},\quad \tilde C_0=-\frac{7 \sqrt{10}+ 5184\alpha_1}{10368},\nn\\
& \beta_1=\frac1{6\sqrt{10}},\quad \beta_2=\frac1{6\sqrt{10}},\quad \alpha_0=-\frac{4\sqrt{10}+ 3240\alpha_1}{135L^2},
\end{align}
and then write
\begin{align}\label{EqHmn2}
\bigg(\square-\frac{32}{L^2}\bigg)H^4_{\mu\nu}&+g_{\mu\nu}\bigg(\frac{4\sqrt{10}+ 3240\alpha_1}{135L^2} \Psi^2\Psi^2+\alpha_1\nabla_\rho\Psi^2\nabla^\rho\Psi^2\bigg)\nn\\
&+\frac1{6\sqrt{10}}\Big(\nabla_{\mu}\Psi^2\nabla_{\nu}\Psi^2
+\Psi^2\nabla_{\mu}\nabla_{\nu}\Psi^2\Big)=0.
\end{align} 
The asymptotic expansion of the LLM solution satisfies this equation up to quadratic order in the mass parameter, independent of the value of the constant $\alpha_1$. Since this constant plays no physical role, we can set it to zero and write 
\begin{align}\label{fredhmn2}
H^4_{\mu\nu}=\check h_{\mu\nu}^{4}+\frac{1}{51840\sqrt{10}}\bigg[&-10g_{\mu\nu}\Big(35\check\psi^2\check\psi^2-\frac{2L^2}{3}\nabla_\rho\check\psi^2\nabla^\rho\check\psi^2-\frac{L^4}{32}\nabla_\rho\nabla_\sigma\check\psi^2\nabla^\rho\nabla^\sigma\check\psi^2\Big)\nn\\
&+113L^2\nabla_{\mu}\check\psi^2\nabla_{\nu}\check\psi^2+103L^2\check\psi^2\nabla_{\mu}\nabla_{\nu}\check\psi^2\\
&+\frac{5L^4}{12}\nabla_{\mu}\nabla^\rho\check\psi^2\nabla_{\nu}\nabla_\rho\check\psi^2-\frac{5L^4}{6}\nabla^\rho\check\psi^2\nabla_{\mu}\nabla_{\nu}\nabla_\rho\check\psi^2\nn\\
&+\frac{L^6}{32}\nabla_{\mu}\nabla^\rho\nabla^\sigma\check\psi^2\nabla_{\nu}\nabla_\rho\nabla_\sigma\check\psi^2+\frac{L^6}{32}\nabla^\rho\nabla^\sigma\check\psi^2\nabla_{\mu}\nabla_{\nu}\nabla_\rho\nabla_\sigma\check\psi^2\bigg].\nn
\end{align}

The equation of motion for the fourth traceless KK graviton mode is the traceless part of \eqref {EqHmn2} and is given by
\begin{align}\label{EqHmn3}
\Big(\square-\frac{32}{L^2}\Big)H^4_{(\mu\nu)}+\frac1{6\sqrt{10}}\Big(\nabla_{(\mu}\Psi^2\nabla_{\nu)}\Psi^2+\Psi^2\nabla_{(\mu}\nabla_{\nu)}\Psi^2\Big)=0,
\end{align}
where
\begin{align}\label{fredhmn2a}
H^4_{(\mu\nu)}=&H^4_{\mu\nu}-\frac14 g_{\mu\nu}g^{\rho\sigma}H^4_{\rho\sigma}\nn\\
=&\check h_{(\mu\nu)}^{4}+\frac1{1080\sqrt{10}}\bigg[\frac{1}{72}g_{\mu\nu}\Big(303\check\psi^2\check\psi^2-\frac{389L^2}{8}\nabla_\rho\check\psi^2\nabla^\rho\check\psi^2+\frac{5L^4}{16}\nabla_\rho\nabla_\sigma\check\psi^2\nabla^\rho\nabla^\sigma\check\psi^2\nn\\
&-\frac{3L^6}{256}\nabla_\rho\nabla_\sigma\nabla_\lambda\check\psi^2\nabla^\rho\nabla^\sigma\nabla^\lambda\check\psi^2\Big)+113L^2\nabla_{\mu}\check\psi^2\nabla_{\nu}\check\psi^2+103L^2\check\psi^2\nabla_{\mu}\nabla_{\nu}\check\psi^2\nn\\
&+\frac{5L^4}{12}\nabla_{\mu}\nabla^\rho\check\psi^2\nabla_{\nu}\nabla_\rho\check\psi^2-\frac{5L^4}{6}\nabla^\rho\check\psi^2\nabla_{\mu}\nabla_{\nu}\nabla_\rho\check\psi^2\nn\\
&+\frac{L^6}{32}\nabla_{\mu}\nabla^\rho\nabla^\sigma\check\psi^2\nabla_{\nu}\nabla_\rho\nabla_\sigma\check\psi^2+\frac{L^6}{32}\nabla^\rho\nabla^\sigma\check\psi^2\nabla_{\mu}\nabla_{\nu}\nabla_\rho\nabla_\sigma\check\psi^2\bigg].
\end{align} 
The last equation is the KK map for the fourth KK graviton mode in quadratic order in the mass parameter. 

\section{Gauge Invariant Operators and Vevs in mABJM Theory}
In the previous section, we defined the physical modes in 4-dimensions using various non-linear KK maps including higher derivatives. These physical modes have corresponding operators by the gauge/gravity dictionary. In this section, we discuss possible operators with conformal dimension $\Delta = 2$ in the ABJM theory and read the $vevs$ of those operators in the large $N$ limit from the vacua of the mABJM theory.

\subsection{Vacua in the mABJM theory}
The mass term in the mABJM theory breaks the SU(4) global symmetry of the ABJM theory to ${\rm SU(2)}\times {\rm SU(2)}\times {\rm U}(1)$. According to the reduced global symmetry, we split the four-complex scalar fields in the ABJM theory as $Y^A=(Z^a,W^{\dagger a})$, where $A=1,2,3,4$ and $a,b=1,2$. Accordingly, the vacuum equation in the mABJM theory is written as
\begin{align}\label{Vac-eq}
&Z^aZ^\dagger_bZ^b-Z^bZ^\dagger_bZ^a=-\frac{\mu k}{2\pi} Z^a,\qquad W^{\dagger a}W_b W^{\dagger b}
-W^{\dagger b} W_b W^{\dagger a}=\frac{\mu k}{2\pi} W^{\dagger a},\nn\\
&W_aZ^bW_b-W_bZ^bW_a=0,\qquad Z^bW_bZ^a-Z^aW_bZ^b=0,
\end{align}
where $\mu$ is a mass parameter. The solutions of those vacuum equations have been obtained in~\cite{Gomis:2008vc} and are presented by a direct sums of two types of irreducible $n\times (n+1)$ matrices ${\cal M}_a^{(n)}~(a=1,2)$ and their Hermitian conjugates, $\bar{\cal M}_a^{(n)}$. These rectangular matrices are referred as the GRVV matrices,
\begin{align}\label{mat-1}
{\cal M}_1^{(n)}&=\left(\begin{array}{cccccc}
\sqrt{n\!}\!\!\!&0&&&&\\&\!\sqrt{n\!-\!1} \!\!&\!0&&&\\
&&\ddots&\ddots&&\\&&&\sqrt{2}&0&\\&&&&1&0\end{array}\right),
\qquad
{\cal M}_2^{(n)}&=
\left(\begin{array}{cccccc}0&1&&&&\\&0&\sqrt{2}&&&\\ &&\ddots&\ddots&&\\
&&&0\!&\!\!\sqrt{n\!-\!1}\!&\\&&&&0&\!\!\!\sqrt{n\!}\end{array}
\right),
\end{align}
where $n = 0, 1,\cdots, N-1$.
The vacuum solutions are given by 
\begin{align}\label{ZW-vacua}
Z^a_0&=\sqrt{\frac{\mu k}{2\pi}}\left(\begin{array}{c}
\begin{array}{cccccc}\mathcal{M}_a^{(n_1)}\!\!&&&&&\\&\!\!\ddots\!&&&&\\
&&\!\!\mathcal{M}_a^{(n_i)}&&& \\ &&& {\bf 0}_{(n_{i+1}+1)\times n_{i+1}}
&&\\&&&&\ddots&\\&&&&&{\bf 0}_{(n_f+1)\times n_f}\end{array}\\
\end{array}\right),\nonumber
\end{align}
\begin{align}
W^{\dagger a}_0&=\sqrt{\frac{\mu k}{2\pi}}\left(\begin{array}{c}
\begin{array}{cccccc}{\bf 0}_{n_1\times (n_1+1)}&&&&&\\&\ddots&&&&\\
&&{\bf 0}_{n_i\times(n_i + 1)} &&&\\
&&& \bar{\mathcal M}_a^{(n_{i+1})}\!\!&&\\&&&&\!\!\ddots\!&\\
&&&&&\!\!\bar{\mathcal M}_a^{(n_f)}\end{array}\\
\end{array}\right).
\end{align}
A given vacuum solution contains $N_n$ rectangular matrices of the type ${\cal M}_a^{(n)}$ and $N_n'$ rectangular matrices of the type $\bar{\cal M}_a^{(n)}$.
The set of parameters $\{N_n,N_n'\}$ completely specifies a vacuum solution and they are called occupation numbers~\cite{Kim:2010mr,Cheon:2011gv}. Since $Z^a$ and $W^{\dagger a}$ are $N\times N$ matrices, the occupation numbers should satisfy the two constraints,
\begin{align}\label{NnNnp}
N = \sum_{n=0}^{N-1}\Big[\left(n+\frac12\right)\left(N_n+ N_n'\right)\Big], 
\qquad 
 \sum_{n=0}^{\infty}N_n = \sum_{n=0}^{\infty}N_n'. 
\end{align}
At quantum level, some of vacuum solutions are not supersymmetric and only a subset of these classical solutions satisfying the conditions, $0\le N_n$ and $N_n'\le k$, remain to be supersymmetric~\cite{Kim:2010mr}.

\subsection{Gauge invariant operators in the ABJM theory}
In general, the CPOs of conformal dimension $\Delta$ in the mABJM theory are given by a trace of products of the four complex scalar fields $Y^A$ and their hermitian conjugates $Y^\dagger_A$,
\begin{align}\label{CPOD}
{\cal O}^{(\Delta)}=C_{A_1\cdots A_{\Delta}}^{B_1\cdots B_{\Delta}}{\rm Tr}\big(Y^{A_1}Y^\dagger_{B_1}\cdots Y^{A_{\Delta}}Y^\dagger_{B_{\Delta}}\big).
\end{align}
These CPOs are dual to the KK scalar modes $\Psi^{I_1}$ with mass $M^2_{\Psi^{I_1}}=\frac{I_1(I_1-6)}{L^2}$ and conformal dimensions $\Delta=\frac{I_1}2,~\{I_1=2,4,6,\cdots\}$~\cite{Jang:2016tbk}. The dual gauge invariant operators for the other KK towers of scalar modes are the descendent of these CPOs, which are obtained by applying the supersymmetry generators of the ${\cal N}=6$ mABJM theory to ${\cal O}^{(\Delta)}$. In particular, the gauge invariant operators dual to the scalar modes $\Phi^{I_1}$ are obtained by applying six supersymmetry generators to the CPO and thus they are given by
\begin{align}\label{GIO}
&{\cal O}_6^{(\Delta)}=C_{a_1a_2a_3A_1\cdots A_{\Delta-6}}^{(6)b_1b_2b_3B_1\cdots B_{\Delta-6}}{\rm STr}\big(\psi^{\dagger a_1}\psi_{b_1}\psi^{\dagger a_2}\psi_{b_2}\psi^{\dagger a_3}\psi_{b_3}Y^{A_1}Y^\dagger_{B_1}\cdots Y^{A_{\Delta-6}}Y^\dagger_{B_{\Delta-6}}\big),
\end{align}
where $\psi^{\dagger a}$ with $a=1,2,3,4$ are the four complex fermionic fields of the ABJM theory and STr denotes symmetrized trace. According to the relations between the mass of the scalar fields and the conformal dimension of the dual operators listed in \cite{Jang:2016tbk}, the masses of the KK scalar modes $\Phi^{I_1}$ are $M^2_{\Phi^{I_1}}=\frac{(I_{1}+12)(I_{1}+6)}{L^2}$ and their conformal dimensions are $\Delta=\frac{I_{1}+12}2,~\{I_1=0,2,4,\cdots\}$. Therefore, the gauge invariant operator dual to the scalar mode $\Phi^4$ is
\begin{align}\label{GIO-1}
&{\cal O}_6^{(\Delta=8)}=C_{a_1a_2a_3A_1A_{2}}^{(6)b_1b_2b_3B_1B_{2}}{\rm STr}\big(\psi^{\dagger a_1}\psi_{b_1}\psi^{\dagger a_2}\psi_{b_2}\psi^{\dagger a_3}\psi_{b_3}Y^{A_1}Y^\dagger_{B_1}Y^{A_{2}}Y^\dagger_{B_{2}}\big),
\end{align}
whereas the scalar field $\Psi^4$ is dual to the CPO,
\begin{align}\label{CPODaa}
{\cal O}^{(\Delta=2)}=C_{A_1A_{2}}^{B_1B_{2}}{\rm Tr}\big(Y^{A_1}Y^\dagger_{B_1} Y^{A_{2}}Y^\dagger_{B_{2}}\big).
\end{align}

%\subsection{CPO with the global symmetry SU(2)$\times$SU(2)$%\times$U(1)}

In our previous paper, we defined the CPO with $\Delta=1$, which reflects the global SU(2)$\times$SU(2)$\times$U(1) symmetry of the mABJM theory. The form of the CPO is given by 
\begin{align}\label{CPO1}
{\cal O}^{(\Delta=1)} = {\cal N}_1{\rm Tr}\left(Y^1Y_1^\dagger + Y^2Y_2^\dagger - Y^3Y_3^\dagger-Y^4Y_4^\dagger\right),
\end{align}
where ${\cal N}_1$ is the normalization factor. The procedure to determine the form of the ${\cal O}^{(\Delta=1)}$ was explained in the Appendix A.4 of~\cite{Jang:2016tbk}. However, we fix the normalization factor ${\cal N}_1$ in a different way, which matches the GKP-W relation~\cite{Gubser:1998bc,Witten:1998qj} in the gauge/gravity dictionary. We will explain the details later.

In this section, we consider the CPO with $\Delta =2$, which reflects the global SU(2)$\times$SU(2)$\times$U(1) symmetry of the mABJM theory. 
Using a similar procedure as in the Appendix A.4 of~\cite{Jang:2016tbk}, we determine the relations among the constants $C_{A_1A_{2}}^{B_1B_{2}}$ in \eqref{CPODaa} and construct the CPO with $\Delta =2$ with the global SU(2)$\times$SU(2)$\times$U(1) symmetry as\footnote{See also the Appendix of the current paper for the details.} 
\begin{align}\label{CPO2}
{\cal{O}}^{(\Delta=2)}=&{\cal N}_2
\left[\sum_{A,B=1}^{2}{\rm Tr}(Y^{A}Y^{\dagger}_{A}Y^{B}Y^{\dagger}_{B})
+\sum_{A,B=1}^{2}{\rm Tr}(Y^{A}Y^{\dagger}_{B}Y^{B}Y^{\dagger}_{A})\right.
\nn\\
&~~~~~~~+\sum_{A,B=3}^{4}{\rm Tr}(Y^{A}Y^{\dagger}_{A}Y^{B}Y^{\dagger}_{B})
+\sum_{A,B=3}^{4}{\rm Tr}(Y^{A}Y^{\dagger}_{B}Y^{B}Y^{\dagger}_{A})\nn\\
&~~~~~~~\left.-3\sum_{A=1}^{2}\sum_{B=3}^{4}{\rm Tr}(Y^{A}Y^{\dagger}_{A}Y^{B}Y^{\dagger}_{B})-3\sum_{A=1}^{2}\sum_{B=3}^{4}{\rm Tr}(Y^{A}Y^{\dagger}_{B}Y^{B}Y^{\dagger}_{A})\right].
\end{align}
where ${\cal N}_2$ is the normalization factor. We will fix the normalization factor later by use of the GKP-W relation.

In order to obtain the $vevs$ of the above CPOs, we expand the complex scalar fields near the vacuum as
\begin{align}\label{Yexp}
Y^A = Y_0^A + \hat Y^A,
\end{align}
where $Y_0^A$'s denote the discrete Higgs vacua discussed above and $\hat Y^A$'s are the complex scalar operators representing fluctuations around the vacua. 
Then the $vev$ of a CPO in the mABJM theory is given by~\cite{Jang:2016tbk}
\begin{align}\label{vevcal}
\langle {\cal O}^{(\Delta)}\rangle_m = {\cal O}^{(\Delta)}\big |_{Y^A=Y^A_0} + \sum_i \langle \delta {\cal O}_i^{(\Delta)}\rangle_0 +{\cal{O}}\Big(\frac1N\Big), 
\end{align}
where $\langle\cdots \rangle_m$ and $\langle\cdots \rangle_0$ denote the $vevs$ of an operator in the mABJM theory and the ABJM theory, respectively, and $\delta {\cal O}_i^{(\Delta)}$ is an operator containing at least one $\hat Y^A$ or $\hat Y^{\dagger A}$. The $\frac1N$-corrections come from the contributions of multi-trace terms. The second term is a one point function in a conformally symmetric ABJM theory and is vanishing. Therefore, in the large $N$ limit, we have 
\begin{align}\label{vevca2}
\langle {\cal O}^{(\Delta)}\rangle_m = {\cal O}^{(\Delta)}\big |_{Y^A=Y^A_0}.
\end{align}
We will display the explicit forms of the $vevs$ for CPOs of conformal dimensions $\Delta = 1$ and $\Delta = 2$ in section 5. 

\section{Asymptotic Behavior of LLM Geometries and 4-dimensional KK Modes}
The metric for the LLM geometries with ${\mathbb Z}_k$ orbifold, which have SO(2,1)$\times {\rm SO}(4)/{\mathbb Z}_k \times {\rm SO}(4)/{\mathbb Z}_k$ isometry~\cite{Auzzi:2009es,Cheon:2011gv}, is given by
 \begin{align}\label{dsFG}
ds^2 =&\frac{ L^2}{4 z^2}\left[dz^2 +\frac{4z^2}{L^2} \big(1+{\tilde g}_1(z,\tau )\big)\left( -dt^2+dw_1^2+dw_2^2\right) \right]\nn\\
&+ \big(1+{\tilde g}_2(z,\tau ) \big)d\tau^2+ \big(1+{\tilde g}_3(z,\tau )\big)ds_{S^3/{\mathbb Z}_k}^2 + \big(1+{\tilde g}_4(z,\tau )\big)ds_{\tilde S^3/{\mathbb Z}_k}^2,
\end{align}
where the $\tilde g_i(z,\tau)$ represent the deviation of the LLM metric from the AdS$_4\times S^7$ background. See \cite{Jang:2016tbk} for details. Similarly, the 4-form field strength of the LLM geometries can be split into the background and the fluctuations. The values of the various KK modes ($h^{I_1}_{\mu\nu}, \phi^{I_1}, \cdots$), introduced in section 2, are read from the asymptotic expansion of $\tilde g_i(z,\tau)$ and the similar functions in 4-form field strength.
In \cite{Jang:2016tbk}, we have listed the full result for all the KK modes up to $\mu_0^2$ order. As mentioned in the previous section, here we focus on the equations of motion for the fourth KK scalar and graviton modes. For the quadratic parts in the equations of motion and in the KK maps discussed in the previous section, we also need the asymptotic expansion of $\check\psi^{2}$. Then we take the following results for the 11-dimensional modes from \cite{Jang:2016tbk}
\begin{align}\label{hhppp}
&\check h_{ij}^4 = \left[-\frac{3L^2\mu_0^2}{4\sqrt{10}}\beta_3^2 + {\cal O}\left(\mu_0^4\right)\right]\eta_{ij},\qquad
\check h_{zz}^4 = -\frac{L^2\mu_0^2}{4\sqrt{10}}\beta_3^2 + {\cal O}\left(\mu_0^4\right), 
\nn \\ 
&\check\psi^4 = -\frac{2\sqrt{10}(\mu_0 z)^2}{35}\left(3780\beta_2^3+758\beta_3^2 -945\beta_2\beta_4 \right) + {\cal O}\left(\mu_0^4\right), 
\nn \\
&\check\phi^4 = -\frac{44\sqrt{10}(\mu_0 z)^2}{7} \beta_3^2+ {\cal O}\left(\mu_0^4\right),
\qquad \check\psi^{2} = -24\beta_3\mu_0 z+{\cal O}(\mu_0^3),
\end{align}
where $\eta_{ij} = {\rm diag}(-1,1,1)$ and
\begin{align}\label{beta3}
\beta_2=C_2-C_1^2,\quad\beta_3 =C_3 - 3 C_1 C_2 +2 C_1^3 ,\quad \beta_4=C_4+3C_2^2-4C_1C_3. 
\end{align}
The parameters $C_p$ were introduced in~\cite{Kim:2014yca,Kim:2016dzw},
\begin{align}\label{Cp}
C_p = \sum_{i=1}^{\infty}(-1)^{i+1} \left( \frac{\tilde x_i}{2\pi l_{{\rm P}}^3 \mu_{0}\sqrt{A}}\right)^p,
\end{align}
where $A$ is defined by
\begin{align}\label{Area}
A = k N -\frac12 \sum_{n=0}^{\infty} \left[l_n (k-l_n) + l_n' (k-l_n')\right]
\end{align}
with the discrete torsions $(l_n,l_n')$ introduced in \cite{Cheon:2011gv}. In the Young diagram representation of the LLM solutions, $A$ means the area of the Young diagram~\cite{Jang:2016tbk}.

In the previous section, we have established the KK maps which relate the above 11-dimensional KK modes to the corresponding canonical 4-dimensional gravity fields. These maps are given in \eqref{fieldRed} and \eqref{fredhmn2a}. These maps express the asymptotic expansions of the fourth KK scalar and graviton modes as follows 
\begin{align}\label{astExp1}
&\Phi^4={\cal O}\Big((\mu_0z)^4\Big),\qquad\Psi^4=-54\sqrt{10}(4\beta_2^3+\beta_3^2-\beta_2\beta_4)(\mu_0z)^2+{\cal O}\Big((\mu_0z)^4\Big)\nn\\
&H^4_{(ij)}=\frac{L^2}{4z^2}\Bigg[-\frac{4(\mu_0z)^2}{\sqrt{10}}\beta_3^2+{\cal O}\Big((z\mu_0)^4\Big)\Bigg]\eta_{ij},\qquad H^4_{(zz)}=\frac{L^2}{4z^2}\Bigg[\frac{12(\mu_0z)^2}{\sqrt{10}}\beta_3^2+{\cal O}\Big((\mu_0z)^4\Big)\Bigg].\end{align}
For clarity of presentation, we also rewrite the similar results for the zeroth and second KK graviton modes obtained in~\cite{Jang:2017gwd} and ~\cite{Jang:2016tbk}, respectively,
\begin{align}\label{Hmn}
&H^0_{ij} = \frac{L^2}{4z^2}\left[-\frac{(\mu_0z)^2}{45}\left( 30 + \beta_3^2\right)+{\cal O} \Big((\mu_0z)^4\Big)\right]\eta_{ij} , \nn\\
&H^0_{zz} = \frac{L^2}{4z^2}\left[- \frac{(\mu_0z)^2}{360}\left(960 + 29\beta_3^2\right) + {\cal O}\Big((\mu_0z)^4\Big)\right],
\nn\\
& H^{2}_{\mu\nu}= \frac{L^2}{4z^2}\left[0+{\cal O}\Big((\mu_0z)^3\Big)\right].\end{align}

The Fefferman-Graham (FG) coordinate system is more convenient for the implementation of the gauge/gravity dictionary. Therefore, we write the asymptotically AdS$_4$ 4-dimensional metric $\big(\hat g_{\mu\nu}=g^{AdS_4}_{\mu\nu}+H^0_{\mu\nu}\big)$ in the FG coordinate by using the coordinate transformation $z =\tilde z + \frac{\mu_0^2 (960+29\beta_3^2)}{1440} \tilde z^3, $
\begin{align}\label{newFG}
ds^2 = \frac{L^2}{4\tilde z^2}\left[ d\tilde z^2 + \left(1- \Big(2 +\frac{\beta_3^2}{16}\Big)(\mu_0 \tilde z)^2 + {\cal O}\Big((\mu_0\tilde z)^4\Big)\right)\eta_{ij}dx^i dx^j\right].
\end{align}
Since all the terms in \eqref{astExp1} are already at least quadratic in $\mu_0$, the above coordinate transformation only amounts to replacing $z$ by $\tilde z$ in those terms. 

As mentioned in the previous section, the scalar field $\Psi^4$ with $M^2_{\Psi^4}=\frac{I(I-6)}{L^2}\big|_{I=4}$ is dual to a CPO of conformal dimension $\Delta=\frac I2\big|_{I=4}=2$ while the scalar field $\Phi^4$ with $M^2_{\Phi^4}=\frac{(I+12)(I+6)}{L^2}\big|_{I=4}$ is dual to a gauge invariant operator with conformal dimension $\Delta= \frac{I+12}2\big|_{I=4}=8$. The GKP-W relation states that the $vev$ of a CPO (${\cal O}^\Delta$) of conformal dimension $\Delta$ is determined by the coefficient of $z^\Delta$ in the asymptotic expansion of the dual scalar field. Thus the $vev$ of the CPO in terms of the holographic renormalizaton~\cite{Balasubramanian:1999re,deHaro:2000vlm,Skenderis:2000in,Bianchi:2001kw,Henningson:1998gx,deBoer:1999tgo,Kraus:1999di,Bianchi:2001de,Martelli:2002sp,Skenderis:2002wp} is given by 
\begin{align}
\langle{\cal O}^{\Delta=2}\rangle_{{\rm HR}}=-54{\mathbb N}\sqrt{10}\mu_0^2(4\beta_2^3+\beta_3^2-\beta_2\beta_4),
\end{align}
where $\mathbb N$ is some normalization factor to be fixed later.

Similarly, the gauge/gravity dictionary maps the metric to the stress-energy tensor $T_{ij}$ of the dual gauge theory. Writing the ($d+1$)-dimensional metric in the FG coordinate
\begin{align}
ds^2 = \frac{L_{\rm AdS}^2}{\tilde z^2}\left[d\tilde z^2 + g_{ij}(x,\tilde z) dx^idx^j\right]
\end{align}
with the asymptotic expansion of the function $g_{ij}(x,\tilde z) $ given by
\begin{align}
g_{ij}(x,\tilde z) = g_{(0)ij}(x) + \tilde z^2 g_{(2)ij}(x) +\cdots + \tilde z^d g_{(d)ij}(x)+\cdots,
\end{align}
then the $vev$ of the stress-energy tensor is given by~\cite{Balasubramanian:1999re,deHaro:2000vlm,Skenderis:2000in,Bianchi:2001kw}
\begin{align}\label{vevTij}
\langle T_{ij}\rangle_{{\rm HR}} = \frac{d L_{{\rm AdS}}^{d-1}}{16\pi G_N}\, g_{(d)ij}. 
\end{align}
From \eqref{newFG} we read that the asymptotic expansion does not contain the $\tilde z^{3}$ term with $d=3$ in \eqref{vevTij}, which implies that the $vev$ of the stress-energy tensor of the mABJM theory is vanishing as required by the supersymmetry of the theory.

The non-zero KK graviton modes $H^2_{\mu\nu}$ and $H^4_{\mu\nu}$ are dual to the operators 
\begin{align}\label{TijYY}
T_{ij}^{(2)}=C_{A}^{B}{\rm STr}\big(T_{ij}Y^{A}Y^\dagger_{B}\big),\qquad 
T_{ij}^{(4)}=C_{AA'}^{BB'}{\rm STr}\big(T_{ij}Y^{A}Y^\dagger_{B}Y^{A'}Y^\dagger_{B'}\big),
\end{align}
respectively. The $vevs$ of these operators are given by 
\begin{align}\label{vevTij1}
\langle T^{(2)}_{ij}\rangle_{{\rm HR}} = {\mathbb N}_{2}\, g^{(2)}_{(d-1)ij},\qquad 
\langle T^{(4)}_{ij}\rangle_{{\rm HR}} = {\mathbb N}_{4}\, g^{(4)}_{(d)ij},
\end{align}
where $g^{(2)}_{(d-1)ij}$ is the coefficient of $\tilde z^{d-1=2}$ in the expansion of $H^2_{ij}$ and $g^{(4)}_{(d)ij}$ is the coefficient of $\tilde z^{d=3}$ in the expansion of $H^4_{ij}$. From \eqref{astExp1} and \eqref{Hmn}, we see that the expansion of $H^2_{ij}$ contains only odd powers of $\tilde z$ whereas the expansion of $H^4_{ij}$ contains only even powers of $\tilde z$. Therefore, the $vevs$ of both $T_{ij}^{(2)}$ and $T_{ij}^{(4)}$ are vanishing.

\section{Vevs of CPOs and GKP-W Relation}
In our previous work~\cite{Jang:2017gwd}, we have constructed the 4-dimensional gravity action with two scalar fields, $T$ and $\Psi_{(1)}$, after the KK reduction from the 11-dimensional supergravity. The field $T$ is dual to a gauge invariant operator, $\tilde{\cal O}^{(2)}=\tilde C_A^B{\rm Tr}\big(\psi^{\dagger A}\psi_{B}\big)$ with $\Delta=2$ and the field $\Psi_{(1)}$ is dual to the CPO \eqref{CPO1}. 

In this section, we focus on the GKP-W relation for the CPOs with $\Delta=1,2$. For that purpose, we consider the 4-dimensional gravity action with two scalar fields, $\Psi_{(1)}$ and $\Psi_{(2)}$,
\begin{align}\label{4dact}
S&= \frac1{16\pi G_4}\int d^4 x\sqrt{-g} \left(\hat R - 2\Lambda\right) -\sum_{i=1}^2\left[ \frac{A_{\Psi_{(i)}}}{2}\int d^4 x\sqrt{-g}\Big(\partial_\mu \Psi_{(i)}\partial^\mu \Psi_{(i)} + M_{\Psi_{(i)}}^2 \Psi_{(i)}^2\Big)\right]
\nn \\
&= \frac{N^2}{3\sqrt{2\pi^2\lambda} L^2}\int d^4 x\sqrt{-g} \left[\hat R - 2\Lambda -\frac12\sum_{i=1}^2 \Big(\partial_\mu \tilde\Psi_{(i)}\partial^\mu \tilde\Psi_{(i)} + M_{\tilde\Psi_{(i)}}^2 \tilde\Psi_{(i)}^2\Big)\right],
\end{align}
where $\frac1{16\pi G_4} = \frac{N^2}{3\sqrt{2\pi^2\lambda} L^2}$ with the 't Hooft coupling $\lambda = N/k$ in the ABJM theory. In order to obtain the normalization which is consistent with the GKP-W relation in the literature, we rescaled the scalar fields as
\begin{align}\label{piG4A}
\tilde \Psi_{(i)} = \sqrt{16\pi G_4 A_{\Psi_{(i)}}}\, \Psi_{(i)}. 
\end{align} 
Solutions for the rescaled fields are read from the asymptotic expansion of the LLM geometries, 
\begin{align} \label{tPsi1}
\tilde \Psi_{(1)} &= -\frac1{\sqrt{2}}\beta_3\mu_0 z + {\cal O}(\mu_0^3),
\nn \\
\tilde \Psi_{(2)} &= -\frac1{\sqrt{2}}\,(4\beta_2^3+\beta_3^2-\beta_2\beta_4)(\mu_0z)^2+{\cal O} (\mu_0^4),
\end{align} 
where we set the scaling factor in \eqref{piG4A} as $\sqrt{16\pi G_4 A_{\Psi_{(1)}}} = \frac1{24\sqrt{2}}$ by reading the value of $A_{\Psi_{(1)}}$ from the equation of motion of $H^{0}_{\mu\nu}$ at $\mu_0^2$ order obtained in~\cite{Jang:2017gwd}. However, the scaling factor $\sqrt{16\pi G_4 A_{\Psi_{(2)}}}$ in \eqref{piG4A} cannot be fixed without the information for the equation of motion of $H^{0}_{\mu\nu}$ at $\mu_0^4$-order. Since we do not have the equation of motion of $H^0_{\mu\nu}$ up to $\mu_0^4$-order, we choose this scaling factor as $\sqrt{16\pi G_4 A_{\Psi_{(2)}}} = \frac1{108\sqrt{5}}$ for later convenience.

As we mentioned in section 4, the GKP-W relation imply, for odd dimensional QFT, the $vev$ of a gauge invariant operator with conformal dimension $\Delta$ is obtained from the holographic renormalization procedure~\cite{Balasubramanian:1999re,deHaro:2000vlm,Skenderis:2000in,Bianchi:2001kw,Henningson:1998gx,deBoer:1999tgo,Kraus:1999di,Bianchi:2001de,Martelli:2002sp,Skenderis:2002wp} in the large $N$ limit,
\begin{align}\label{vCPOD}
\langle {\cal O}^{(\Delta)} \rangle_{{\rm HR}} = \frac{N^2}{3\sqrt{2\pi^2\lambda}} \left(2\Delta - d\right) \tilde\psi^{(i)}_{\Delta},
\end{align}
where $\tilde\psi^{(i)}_\Delta$ is the coefficient of $z^\Delta$ in the asymptotic expansion of the field $\tilde\Psi_{(i)}$. 
Inserting the solutions \eqref{tPsi1} into \eqref{vCPOD}, we obtain
\begin{align}\label{vCPO12}
\langle {\cal O}^{(\Delta=1)} \rangle_{{\rm HR}} &= -\frac{N^2}{3\sqrt{2\pi^2\lambda}}\,\tilde\psi^{(1)}_1 =\frac{N^2\beta_3\mu_0}{6\pi\sqrt{\lambda}}, 
\nn \\
\langle {\cal O}^{(\Delta=2)} \rangle_{{\rm HR}} &= \frac{N^2}{3\sqrt{2\pi^2\lambda}}\,\tilde\psi^{(2)}_2 =-\frac{N^2}{6\pi\sqrt{\lambda}} \left(4\beta_2^3+\beta_3^2-\beta_2\beta_4\right)\mu_0^2.
\end{align}
The normalization factors ${\cal N}_{1,2}$ of the CPOs defined in \eqref{CPO1} and \eqref{CPO2} are determined from \eqref{vCPO12}. 
For the CPO of conformal dimension $\Delta = 1$, the $vev$ \eqref{vevca2} of the mABJM theory in the large $N$ limit can be read as~\cite{Jang:2016tbk}
\begin{align}\label{vevO1}
 \langle {\cal O}^{(\Delta=1)}\rangle_m={\cal N}_1{\rm Tr}\left(Y^1Y_1^\dagger + Y^2Y_2^\dagger - Y^3Y_3^\dagger-Y^4Y_4^\dagger\right)\Big |_{Y^A= Y_0^A} = \frac{2{\cal N}_1 N^2\beta_3\mu_0}{3\pi \sqrt{\lambda}},
\end{align}
where $\langle\cdots\rangle_m$ represents the $vev$ of an operator in the mABJM theory.
Comparing the $vev$ in terms of the holographic renormalization in \eqref{vCPO12} with that of the mABJM theory in \eqref{vevO1}, we fix the normalization factor of ${\cal O}^{(1)}$ as ${\cal N}_1=\frac14$. Thus the definition of ${\cal O}^{(1)}$ in this paper has a factor of $\frac{1}{\sqrt{2}}$ difference from that of the previous paper~\cite{Jang:2016tbk,Jang:2017gwd}. 

%%%%%%%%%%%%%%%%%%%%%%%%%%
\begin{figure}%[!h]
\centerline{\epsfig{figure=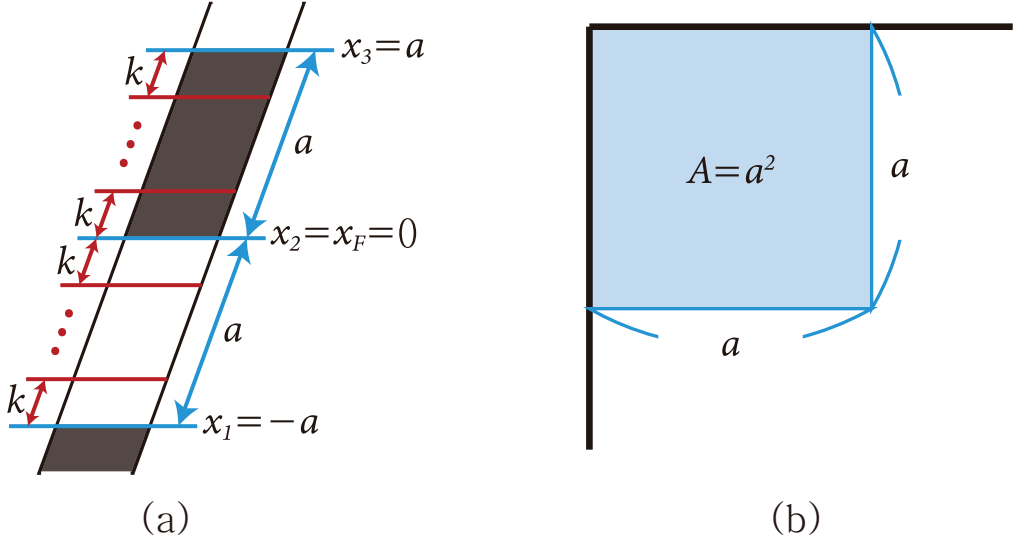,
height=75mm}}
\caption{
\small (a) Symmetric droplet representation of the LLM geometry, where the number of black strip is one, the length of it is $a$, and $k=\frac{a}{n}$ with integer $n$. 
(b) Young diagram corresponding to the droplet picture (a).}
\label{Fig1}
\end{figure}
%%%%%%%%%%%%%%%%%%%%%%%

In order to fix the normalization factor ${\cal N}_2$ in \eqref{CPO2}, we consider a symmetric droplet case with $k\ne 1$. The corresponding droplet and Young diagram representations in the LLM geometries are depicted in Fig. \ref{Fig1}. In this case, we set $k=\frac{a}{n}$, $N= na$, and $A = kN = a^2$. Then by fixing the coordinate of the Fermi level as $\tilde x_2 = \tilde x_F=0$\footnote{For the details of the droplet and Young diagram representations in the LLM geometries, see \cite{Cheon:2011gv,Jang:2016tbk}.}, we obtain 
\begin{align}
C_1 = C_3 = 0, \quad C_2 = C_4 = 2. 
\end{align}
Using these values in the second line of \eqref{vCPO12}, we obtain 
\begin{align}\label{vCPO12b}
\langle {\cal O}^{(\Delta=2)} \rangle_{{\rm HR}} =-\frac{2N^2}{3\pi\sqrt{\lambda}} \mu_0^2.
\end{align}

Now we try to calculate the corresponding {\it vev} in the field theory side. For the symmetric droplet case, one can also assign the discrete torsions as 
\begin{align}
(l_0,l_1, \cdots, l_{n-1}) = \left(\frac{a}{n},\frac{a}{n},\cdots,\frac{a}{n}\right),\qquad 
(l_0',l_1', \cdots ,l_{n-1}') = \left(\frac{a}{n},\frac{a}{n},\cdots,\frac{a}{n}\right).
\end{align}
Other values of discrete torsions are vanishing. 
Identifying the discrete torsions $\{l_n,l_n'\}$ with the occupation numbers of GRVV matrices $\{N_n,N_n'\}$, we calculate the $vev$ of ${\cal O}^{(\Delta=2)}$ in \eqref{CPO2} in the large $N$ limit, 
\begin{align}\label{CPO2-2}
\langle {\cal O}^{(\Delta=2)} \rangle_m= \frac{2k\mu_0 ^2{\cal N}_2 N^2}{\pi^2 } + {\cal O} (N),
\end{align}
where we have used the relations 
\begin{align}
&{\rm Tr}\left(\sum_{A=1}^4 Y^AY_A^\dagger Y^AY_A^\dagger\right) \Bigg|_{Y^A= Y_0^A}= \frac{4k\mu_0^2 N^2}{3\pi^2 }+ {\cal O} (N),
\nn \\
&{\rm Tr}\left(Y^1Y_1^\dagger Y^2Y_2^\dagger + Y_1^\dagger Y^1 Y_2^\dagger Y^2+Y^3Y_3^\dagger Y^4Y_4^\dagger + Y_3^\dagger Y^3 Y_4^\dagger Y^4\right)\bigg|_{Y^A= Y_0^A}
=\frac{2k\mu_0^2 N^2}{3\pi^2 }+{\cal O} (N).
\end{align}
%%%%%%%%%%%%%%%%%%%%%%%%%%
\begin{figure}%[!h]
\centerline{\epsfig{figure=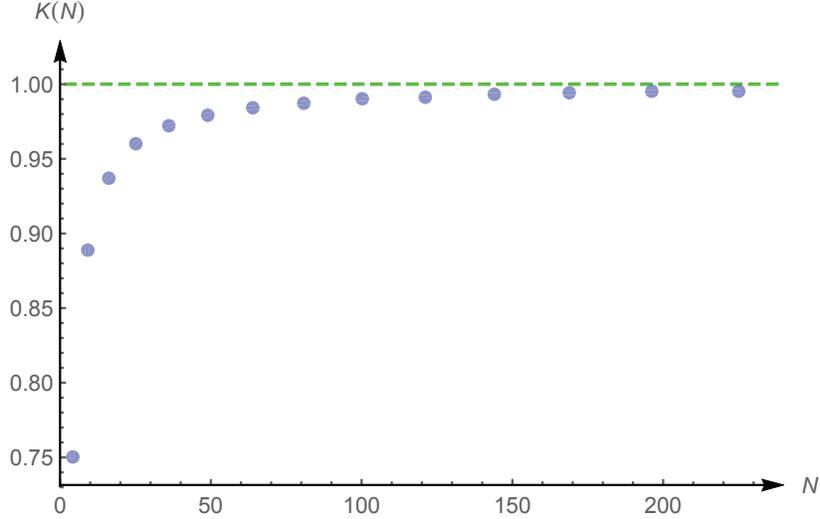,
height=70mm}}
\caption{
\small Validity of the holographic renormalization for the CPO of conformal dimension $\Delta=2$ in the square-shaped Young diagram of the LLM geometries at large $N$. The vertical axis is $ K(N) = \frac{\langle {\cal O}^{(\Delta=2)} \rangle_m }{\langle {\cal O}^{(\Delta=2)} \rangle_{{\rm HR}}}$ with $N=4,9,16,\cdots, 225$.}
\label{Fig2}
\end{figure}
%%%%%%%%%%%%%%%%%%%%%%%
Other combinations of the traces in \eqref{CPO2} are vanishing due to the gauge choice of the vacuum solutions in~\cite{Gomis:2008vc}. Comparing the $vev$ in the field theory side with that in gravity theory side, we fix the normalization factor in \eqref{CPO2} as 
\begin{align}\label{norm2}
{\cal N}_2 = -\frac{\pi}{3\sqrt{kN}}.
\end{align}
We examine validity of the holographic renormalization \eqref{vCPO12b} at large $N$ in Fig. \ref{Fig2}. 

\section{Conclusion}
In this paper, we obtained the $vevs$ of gauge invariant operators up to $\mu_0^2$-order in terms of the holographic renormalization in the mABJM theory. 
We found that the $vevs$ of gauge invariant operators are vanishing up to $\mu_0^2$-order expect for the case of the CPOs with conformal dimension $\Delta = 1,2$. For the latter cases, the $vevs$ were obtained using the KK holography in the large $N$ limit.
In order to show validity of the holographic relation, we compared the $vevs$ from the supersymmetric vacua of the mABJM theory with those from the LLM solutions. Our results for the CPO of conformal dimension $\Delta=2$ are limited to the cases of the LLM solutions, which are represented by a square-shaped Young diagrams. We showed that the $vevs$ obtained from the mABJM theory with an appropriate normalization of the CPO of conformal dimension $\Delta = 2$ approach those obtained from the holographic renormalization at large $N$.

The result we obtained in this paper is a further confirmation of the claim in \cite{Jang:2016tbk} about duality between the mABJM theory and the 11-dimensional supergravity on the LLM geometry. However, in the present case the procedure is highly non-trivial. 
In order to read the $vevs$ of the CPO of conformal dimension $\Delta = 2$ from the asymptotic expansion of the LLM solutions, we need to carry out the KK reduction of the 11-dimensional supergravity and then construct a 4-dimensional gravity on the asymptotic AdS$_{4}$ background. Unlike the case of the CPO of conformal dimension $\Delta=1$, we need to establish the KK maps in the quadratic order between the 4-dimensional fields and the 11-dimensional fields. 
The KK maps include the non-trivial field redefinitions, which are required to absorb higher derivative terms and result in the canonical equations of motion for the 4-dimensional fields. 
Identifying the 4-dimensional fields obtained from the KK maps with the fluctuations obtained from the asymptotic expansion of the LLM solutions, we read the asymptotically AdS$_4$ solutions in the 4-dimensional equations of motion. We read the $vevs$ of the CPO of conformal dimension $\Delta = 2$ from those asymptotic solutions in 4-dimensions. We also confirm that the $vevs$ of other gauge invariant operators which are not CPO as well as those of the massive KK graviton modes are vanishing.

In the previous work~\cite{Jang:2016tbk}, we showed that the $vevs$ of ${\cal O}^{(\Delta=1)}$ for any LLM solutions in the holographic renormalization method are exactly the same as those of the mABJM theory in the large $N$ limit, i.e., $\langle {\cal O}^{(\Delta=1)} \rangle_{{\rm HR}} = \langle {\cal O}^{(\Delta=1)} \rangle_m$. This result heavily depends on the fact that the curvature in the asymptotic limit $(\mu_0 z\ll 1)$ becomes weak for any LLM solutions~\cite{Hyun:2013sf}. 
Since the $vev$ $\langle {\cal O}^{(\Delta=1)} \rangle_{{\rm HR}}$ is completely determined by the asymptotic expansion of the LLM solutions in $\mu_0$-order~\cite{Jang:2016tbk}, one can expect that the relation $\langle {\cal O}^{(\Delta=1)} \rangle_{{\rm HR}} = \langle {\cal O}^{(\Delta=1)} \rangle_m$ in the large $N$ limit is satisfied for all LLM solutions. However, by increasing the $\mu_0 z$-value in the LLM geometry, we notice that some LLM geometries, which include short edges in the Young diagram representation, become strongly curved even in the large $N$ limit~\cite{Hyun:2013sf}. 
Therefore, in order to obtain the correct holographic relation $(\Delta\ne 1$) for LLM geometries including strongly curved regions, one needs quantum corrections from the gravity side in the large $N$ limit, 
\begin{align}\label{mHR}
\langle {\cal O}^{(\Delta)} \rangle_m = \langle {\cal O}^{(\Delta)} \rangle_{{\rm HR}} + {\rm quantum\,\, corrections}.
\end{align}
In other words, the LLM geometries with square-shaped Young diagrams do not include any short edges in the large $N$ limit and thus these geometries are weakly curved over all transverse regions. For these LLM geometries, we expect that the holographic relation \eqref{mHR} is satisfied without quantum corrections in the gravity side. 
In this paper, we examined validity of the $vevs$ of ${\cal O}^{(\Delta=2)}$ in the holographic renormalization for the square-shaped Young-diagrams in the LLM geometries, and showed that $\langle {\cal O}^{(\Delta=2)}\rangle_{{\rm HR}}$ is approaching the value of $\langle {\cal O}^{(\Delta=2)}\rangle_m$ in the field theory side by increasing $N$. This result matches our expectation. It is also intriguing to examine the relation \eqref{mHR} for other Young diagrams in the LLM geometries. 

\section*{Acknowledgements}
OK appreciates APCTP for its hospitality during completion of this work and DT would like to thank the physics department of Addis Ababa University for hospitality, during the visit to present part of this work. This work was supported by the National Research Foundation of Korea(NRF) grant with grant number NRF-2016R1D1A1B03931090 (Y.K.), NRF-2017R1D1A1A09000951 (O.K.), and NRF-2017R1D1A1B03032523 (D.T.).

\appendix 
\section{$C^{I_{1}=4}$ and $C^{(\Delta=2)}$}\label{CabD1}
In this Appendix, we determine the coefficients $C^{I_1=4}_{i_1\cdots i_{4}}$ which define the fourth scalar spherical harmonics on $S^7$ and the coefficients $C^{A_1 A_{2}}_{B_1 B_{2}}$ which defines the CPO of conformal dimension $\Delta = 2$. To that end, we start from the definition of the fourth scalar spherical harmonics on $S^7$,
\begin{align}\label{y4}
Y^{4}
=
\frac{1}{L^{4}}\sum_{i,j,k,l=1}^{8}C_{ijkl}x^{i}x^{j}x^{k}x^{l}
\end{align}
with the ${\mathbb R}^8$ coordinates $x^i$'s which are restricted to $S^7$ as follows,
\begin{align}\label{R8}
&x^1=L\Big(\frac{1+\tau}2\Big)^{\frac12}\cos\big(\frac\theta2\big)\cos\big(\frac{\phi+\psi}2\big),\quad x^2=L\Big(\frac{1+\tau}2\Big)^{\frac12}\cos\big(\frac\theta2\big)\sin\big(\frac{\phi+\psi}2\big),\nn\\
&x^3=-L\Big(\frac{1+\tau}2\Big)^{\frac12}\sin\big(\frac\theta2\big)\sin\big(\frac{\phi-\psi}2\big),\quad
x^4=L\Big(\frac{1+\tau}2\Big)^{\frac12}\sin\big(\frac\theta2\big)\cos\big(\frac{\phi-\psi}2\big),\nn\\
&x^5=L\Big(\frac{1-\tau}2\Big)^{\frac12}\cos\big(\frac{\tilde\theta}2\big)\cos\big(\frac{\tilde\phi+\tilde\psi}2\big),\quad x^6=L\Big(\frac{1-\tau}2\Big)^{\frac12}\cos\big(\frac{\tilde\theta}2\big)\sin\big(\frac{\tilde\phi+\tilde\psi}2\big),\nn\\
&x^7=-L\Big(\frac{1-\tau}2\Big)^{\frac12}\sin\big(\frac{\tilde\theta}2\big)\sin\big(\frac{\tilde\phi-\tilde\psi}2\big),\quad
x^8=L\Big(\frac{1-\tau}2\Big)^{\frac12}\sin\big(\frac{\tilde\theta}2\big)\cos\big(\frac{\tilde\phi-\tilde\psi}2\big).
\end{align}
The coefficients $C_{ijkl}$ are traceless under the contraction of any two indices and also are totally symmetric. Here we are interested in the scalar spherical harmonics on $S^{7}$ with ${\rm SO}(4)\times {\rm SO}(4)$ symmetry,
\begin{align}\label{Y4res}
Y^{4}=\widetilde{\cal N}_4\left(1-5\tau^{2}\right),
\end{align} 
where $\widetilde{\cal N}_4$ is a normalization factor. Subsequently inserting \eqref{R8} into \eqref{y4}, using the tracelessness and the symmetric conditions, and comparing with \eqref{Y4res}, we obtain
\begingroup
\allowdisplaybreaks
\begin{align}
&3C_{1133}=3C_{1144}=C_{3333}=C_{1111}=4\widetilde{\cal N}_4,\nn\\
&3C_{5577}=3C_{5588}=C_{7777}=C_{5555}=4\widetilde{\cal N}_4,\nn\\
&C_{1166}=C_{1177}=C_{1188}=C_{3355}=C_{3366}=C_{3377}=C_{3388}= C_{1155}=-2\widetilde{\cal N}_4,\nn\\
&\textrm{the others = 0}.
\end{align}
\endgroup

In order to determine the coefficients $C^{A_1A_2}_{B_1B_2}$ of the CPO of conformal dimension $\Delta = 2$, we need to rewrite the scalar spherical harmonics in terms of ${\mathbb C}^4$ coordinates $y^{A}=x^{2A-1}+ix^{2A}$ as
\begin{align}\label{Y4}
Y^{4}=\frac{1}{L^{4}}\sum_{A,B,C,D=1}^{4}\widetilde{C}_{ABCD}y^{A}y^{\dagger}_{B}y^{C}y^{\dagger}_{D},
\end{align}
The coefficients $\widetilde{C}_{ABCD}$ satisfy the same conditions as $C_{ijkl}$ and the values of the former are determined from the values of the later as follows
\begingroup
\allowdisplaybreaks
\begin{align}
&\widetilde{C}_{1111}=\frac{3}{8}C_{1111}+\frac{3}{4}C_{1122}+\frac{3}{8}C_{2222}=C_{1111}=4\widetilde{\cal N}_4,\nn\\
&\widetilde{C}_{2222}=\frac{3}{8}C_{3333}+\frac{3}{4}C_{3344}+\frac{3}{8}C_{4444}=C_{1111}=4\widetilde{\cal N}_4,\nn\\
&\widetilde{C}_{3333}=\frac{3}{8}C_{5555}+\frac{3}{4}C_{5566}+\frac{3}{8}C_{6666}=C_{1111}=4\widetilde{\cal N}_4,\nn\\
&\widetilde{C}_{4444}=\frac{3}{8}C_{7777}+\frac{3}{4}C_{7788}+\frac{3}{8}C_{8888}=C_{1111}=4\widetilde{\cal N}_4,\nn\\
&\widetilde{C}_{1122}=\widetilde{C}_{1221}=\frac{3}{4}\left(C_{1133}+C_{1144}
+C_{2233}+C_{2244}\right)=C_{1111}=4\widetilde{\cal N}_4,\nn\\
&\widetilde{C}_{3344}=\widetilde{C}_{3443}=\frac{3}{4}\left(C_{5577}+C_{5588}
+C_{6677}+C_{6688}\right)=C_{1111}=4\widetilde{\cal N}_4,\nn\\
&\widetilde{C}_{1133}=\widetilde{C}_{1441}=\frac{3}{4}\left(C_{1155}+C_{1166}
+C_{2255}+C_{2266}\right)=-\frac{3}{2}C_{1111}=-6\widetilde{\cal N}_4,\nn\\
&\widetilde{C}_{1144}=\widetilde{C}_{1441}=\frac{3}{4}\left(C_{1177}+C_{1188}
+C_{2277}+C_{2288}\right)=-\frac{3}{2}C_{1111}=-6\widetilde{\cal N}_4,\nn\\
&\widetilde{C}_{2233}=\widetilde{C}_{2332}=\frac{3}{4}\left(C_{3355}+C_{3366}
+C_{4455}+C_{4466}\right)=-\frac{3}{2}C_{1111}=-6\widetilde{\cal N}_4,\nn\\
&\widetilde{C}_{2244}=\widetilde{C}_{2442}=\frac{3}{4}\left(C_{3377}+C_{3388}
+C_{4477}+C_{44488}\right)=-\frac{3}{2}C_{1111}=-6\widetilde{\cal N}_4,\nn\\
&\textrm{the others = 0}.
\end{align}
\endgroup
Finally, we identify the coefficients $\widetilde{C}_{ABCD}$ with the coefficients $C^{A_1A_2}_{B_1B_2}$ of the CPO and thus can write 
\begin{align}
{\cal{O}}^{(\Delta=2)}=&\sum_{A,B,C,D=1}^{4}\widetilde{C}_{ABCD}{\rm Tr}(Y^{A}Y^{\dagger_{B}}Y^{C}Y^{\dagger}_{D}),\nn\\
=&{\cal N}_2
\left[\sum_{A,B=1}^{2}{\rm Tr}(Y^{A}Y^{\dagger}_{A}Y^{B}Y^{\dagger}_{B})
+\sum_{A,B=1}^{2}{\rm Tr}(Y^{A}Y^{\dagger}_{B}Y^{B}Y^{\dagger}_{A})\right.
\nn\\
&~~~~~~~+\sum_{A,B=3}^{4}{\rm Tr}(Y^{A}Y^{\dagger}_{A}Y^{B}Y^{\dagger}_{B})
+\sum_{A,B=3}^{4}{\rm Tr}(Y^{A}Y^{\dagger}_{B}Y^{B}Y^{\dagger}_{A})\nn\\
&~~~~~~~\left.-3\sum_{A=1}^{2}\sum_{B=3}^{4}{\rm Tr}(Y^{A}Y^{\dagger}_{A}Y^{B}Y^{\dagger}_{B})-3\sum_{A=1}^{2}\sum_{B=3}^{4}{\rm Tr}(Y^{A}Y^{\dagger}_{B}Y^{B}Y^{\dagger}_{A})\right],
\end{align}
where ${\cal N}_2=2\widetilde{\cal N}_4$.

\end{document}